\documentclass[lettersize,journal]{IEEEtran}
\usepackage{amsmath,amsfonts}

\usepackage{caption}
\usepackage{textcomp}
\usepackage{stfloats}
\usepackage{url}
\usepackage{verbatim}
\usepackage{graphicx}
\usepackage{cite}
\usepackage{multirow,enumitem,adjustbox}
\usepackage{booktabs,threeparttable,makecell}
\usepackage{xcolor}
\usepackage{subcaption}
\usepackage{enumitem}

\usepackage{colortbl}
\usepackage[colorlinks=true,citecolor=green]{hyperref}
\usepackage{cleveref}
\usepackage{rotating}
\usepackage{tablefootnote}
\usepackage[edges]{forest}
\usepackage{tikz}
\usepackage{tikz-qtree}
\usetikzlibrary{arrows.meta, positioning, shapes.geometric}

\usepackage[linesnumbered,ruled,vlined]{algorithm2e}

\SetCommentSty{mycommfont}

\definecolor{lightgray}{RGB}{211, 211, 211}
\definecolor{hiddendraw}{RGB}{205, 44, 36}

\begin{document}

\title{Overview of Speaker Modeling and Its Applications: From the Lens of Deep Speaker Representation Learning}


\author{
Shuai Wang, \IEEEmembership{Member, IEEE,}
Zhengyang Chen, \IEEEmembership{Student Member, IEEE,}
Kong Aik Lee, \IEEEmembership{Senior Member, IEEE,}
Yanmin Qian, \IEEEmembership{Senior Member, IEEE,}
and Haizhou Li, \IEEEmembership{Fellow, IEEE}

{
    \thanks{This work is supported by China NSFC projects under Grants No. 62401377 and 62271432; Shenzhen Science and Technology Program ZDSYS20230626091302006; Shenzhen Science and Technology Research Fund (Fundamental Research Key Project Grant No. JCYJ20220818103001002). Zhengyang Chen and Yanmin Qian are supported in part by China NSFC projects under Grants No. 62122050 and 62071288; in part by Shanghai Municipal Science and Technology Commission Project under Grant 2021SHZDZX0102.}
    \thanks{Shuai Wang and Haizhou Li are with Shenzhen Research Institute of Big Data, School of Data Science, The Chinese University of Hong Kong, Shenzhen, Guangdong 518172, China (e-mail:wangshuai@cuhk.edu.cn; haizhouli@cuhk.edu.cn)}
    \thanks{Zhengyang Chen and Yanmin Qian are with the Auditory Cognition and Computational Acoustics Lab, the Department of Computer Science and Engineering and the MoE Key Laboratory of Artificial Intelligence, AI Institute, Shanghai Jiao Tong University, Shanghai 200240, China (e-mail: zhengyang.chen@sjtu.edu.cn; yanminqian@sjtu.edu.cn).}  
    \thanks{Kong Aik Lee is with the Department of Electrical and Electronic Engineering, The Hong Kong Polytechnic University, Hong Kong (kong-aik.lee@polyu.edu.hk).}

}
}

\markboth{Journal of \LaTeX\ Class Files,~Vol.~14, No.~8, August~2021}%
{Shell \MakeLowercase{\textit{et al.}}: A Sample Article Using IEEEtran.cls for IEEE Journals}


\maketitle

\begin{abstract}

Speaker individuality information is among the most critical elements within speech signals. By thoroughly and accurately modeling this information, it can be utilized in various intelligent speech applications, such as speaker recognition, speaker diarization, speech synthesis, and target speaker extraction. 
In this overview, we present a comprehensive review of neural approaches to speaker representation learning from both theoretical and practical perspectives. Theoretically, we discuss speaker encoders ranging from supervised to self-supervised learning algorithms, standalone models to large pretrained models, pure speaker embedding learning to joint optimization with downstream tasks, and efforts toward interpretability. Practically, we systematically examine approaches for robustness and effectiveness, introduce and compare various open-source toolkits in the field. Through the systematic and comprehensive review of the relevant literature, research activities, and resources, we provide a clear reference for researchers in the speaker characterization and modeling field, as well as for those who wish to apply speaker modeling techniques to specific downstream tasks.

\end{abstract}


\section{Introduction}
\label{sec:intro}
\begin{figure*}[ht!]
  \centering
  \includegraphics[width=0.86\textwidth]{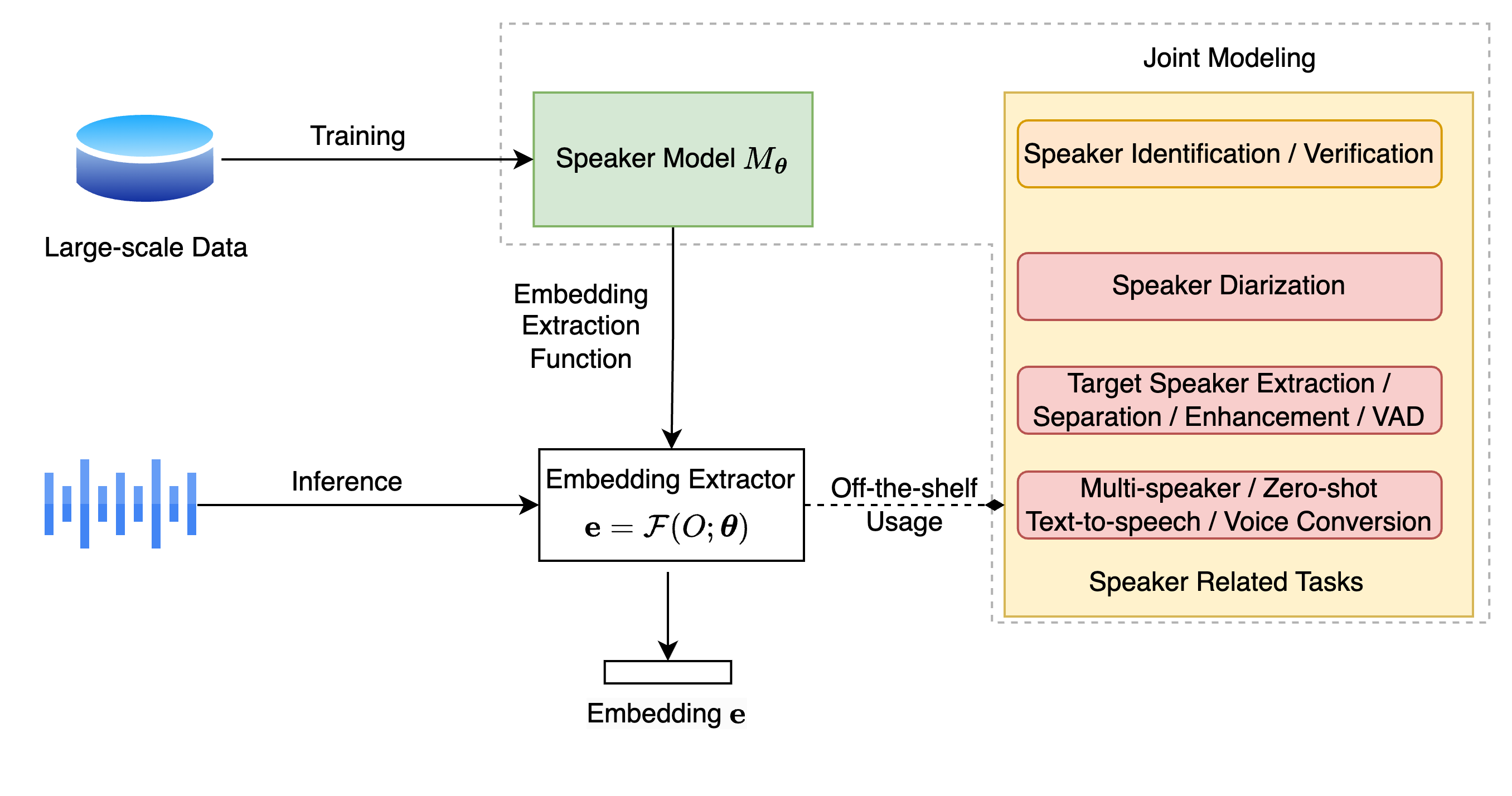}
    \caption{Typical Flow of Speaker Representation Learning and its applications. The speaker model $M_{\boldsymbol{\theta}}$ can be trained in advance and applied to related tasks in an off-the-shelf manner, or joint optimized with the target task.}
  \label{fig:overall}
\end{figure*}

\IEEEPARstart{S}{peaker} modeling aims to represent and recognize the unique characteristics of an individual by analyzing voice patterns or attributes embedded within speech signals. It is a critical technology in the field of speech processing, with wide-ranging applications. This technology encompasses essential details pertaining to the speaker's individuality and vocal characteristics. Most significantly, it underpins the crucial task of speaker recognition, which enables the direct identification of an individual based on their voice. This capability is extensively leveraged in diverse fields, such as biometric authentication~\cite{singh2018voice}, surveillance~\cite{kiktova2015speaker}, personalized services~\cite{mcgraw2016personalized}, and forensic scenarios~\cite{campbell2009forensic}.

Beyond tasks directly related to speaker information, such as speaker recognition~\cite{bai2021speaker, speaker_survey1, speaker_survey2, speaker_survey3, speaker_survey4} and speaker diarization~\cite{diarization_survey1, diarization_survey2}, the applications of speaker modeling extend into speech generation tasks such as voice cloning~\cite{arik2018neural}, speech synthesis~\cite{ttsjia2018transfer}, and voice conversion~\cite{liu2021non}. By learning and mimicking a speaker's voice characteristics, it supports high-quality personalized speech generation, which is extensively used in virtual assistants, gaming characters, and more, thereby enhancing interactive user experiences. Additionally, by transforming one speaker's voice into the characteristics of another, usually a non-existent pseudo speaker, it enables speaker anonymization~\cite{fang2019speaker,tomashenko2024voiceprivacy}, ensuring personal privacy and data security.
Recently, speaker-dependent front-end processing algorithms have garnered significant attentions. These include technologies such as target speaker voice activity detection~\cite{medennikov2020target}, target speaker enhancement~\cite{ji2020speaker}, and target speaker extraction~\cite{zmolikova2023neural}. By incorporating additional speaker information, these traditional front-end processing algorithms can more effectively focus on a specific individual, thereby enhancing the quality and clarity of speech signals.
Overall, precise modeling and a comprehensive understanding of speaker information can significantly contribute to richer and more intelligent interaction experiences within speech processing systems.

Speaker modeling has been studied for many years, primarily through the task of speaker recognition. Its performance has been improved continuously in the past decades with many new techniques. Some of the earliest approaches can be traced back to work at Bell Laboratories during World War II using then newly developed ``\textit{sound spectrograph}''~\cite{koenig1946sound}. Approaches to making speaker recognition fully automatic were first investigated starting in the 1960s using spectrogram template matching (with the use of the term “voiceprint” as an analogy to a fingerprint). Research into new methods for automatic speaker recognition continued thereafter as computer processing and storage increased. Since the 1980s, speaker modeling technology has evolved from the initial Vector Quantization (VQ) algorithms \cite{li1983approach, soong1987report}, Hidden Markov Models (HMM) \cite{tisby1991application, savic1990variable}, Artificial Neural Networks (ANN) \cite{carey1991speaker, bennani1991use, farrell1994speaker, wang1997speaker}, to the Gaussian Mixture Model-Universal Background Model (GMM-UBM) framework \cite{reynolds2000speaker}, with each advancement bringing improved performance to speaker recognition. Over the past two decades, we have witnessed four paradigm shifts or technological transitions. Below, we highlight the core components of each technological advancement and refer readers to relevant references for further details.

\vspace{-1em}

\subsection{Paradigm Shift in the Past Two Decades}
\textbf{From VQ to GMM}. Vector quantization (VQ) and Gaussian mixture model (GMM) have been used to model the distribution of voice features. The VQ technique was first used for speech coding~\cite{makhoul1985vector} and speech recognition~\cite{shore1983discrete,rumelhart1985learning}. In speaker recognition, one VQ codebook is constructed (e.g., k-means) for each speaker using acoustic vectors\cite{li1983approach, soong1987report}. The codebook that provides the smallest VQ distortion (measured as the average distance to the centroids) indicates the identified speaker. In the GMM approach, one GMM is trained for each speaker using acoustic vectors. In addition to the mean vectors (correspond to centroids in VQ), a GMM has covariance matrices (uncertainty) and mixtures weights (frequency of occurrence). The GMM that provides the highest log-likelihood indicates the identified speaker. The major advancement from VQ to GMM is uncertainty modeling. Each Gaussian component in the GMM models uncertainty with a covariance matrix, which is absent in VQ.

\textbf{From GMM-EM to GMM-UBM}. Entering the 21st century, the GMM-UBM framework became one of the most representative models. The idea of a universal background model, or UBM, was first conceived in \cite{reynolds2000speaker}. A UBM is a GMM trained using a large number of speech utterances from many speakers. It represents the general voice characteristic of the speaker population in an application (English, male/female, and telephony). Given an enrollment utterance, a speaker model is obtained by adapting the parameters from the UBM (typically the mean vectors while keeping the weights and covariance matrices) via the so-called Maximum a \textit{Posteriori} (MAP) adaptation~\cite{gauvain1994maximum}. This has been shown to provide a significant advantage in terms of memory usage, computation, and well-behaved log-likelihood ratio score compared to that of GMM-EM. In the latter, the GMM parameters are estimated from scratch with random initialization followed by iterative updates using the expectation-maximization (EM) algorithm. GMM-UBM has limitations during the speaker enrollment process, updating only a subset of Gaussian components, whereas an ideal adaptation process would involve a global update of the entire UBM. To address this, Gaussian supervectors (GSV) are formed by concatenating the mean vectors of GMMs adapted from a UBM ~\cite{kenny2003new,campbell2006support}. Subsequently, the introduction of Support Vector Machines (SVM) as a scoring backend led to significant performance improvements, resulting in the well-known GSV-SVM framework~\cite{campbell2006svm}.

\textbf{From supervector to i-vector}. The term `supervector' was used to reflect the fact the size of the vector representation is much larger than the acoustic feature vectors. The limitations of the GSV became apparent in handling complex and varying recording channels and environments. The high-dimensional GSV contained an excess of non-speaker-specific information, imposing significant demands on memory and computational resources. To resolve these issues, researchers proposed a range of channel compensation and dimensionality reduction algorithms, such as Eigen-channel \cite{kenny2005joint}, Eigen-voice \cite{kenny2005eigenvoice}, and Joint Factor Analysis (JFA) \cite{kenny2007joint} methods. The further development of JFA deepened the study in GSV space, indicating the inherent speaker-discriminative capability within channel factors. Building on this, a unified low-dimensional space to jointly model speaker and channel spaces leads to the widely used \textit{i}-vector framework \cite{dehak2011front}. An \textit{i}-vector is the reduced dimensional representation by confining the supervector to a low-rank vector space. \textit{i}-vector was invented around the emergence of the iPhone with Siri, though formally the ``\textit{i}'' stands for identity vector (i.e., the speaker identity). Although \textit{i}-vector had limitations in decoupling speaker and channel information, necessitating additional channel compensation, they effectively improved speaker recognition accuracy through methods such as Probabilistic Linear Discriminant Analysis (PLDA) \cite{ioffe2006probabilistic}.

\textbf{From Generative to discriminative speaker embedding}. Speaker embedding aims to represent variable-length speech utterances as fixed-length vectors with an additional constraint that the embeddings from the same speaker are close to each other, while those from different speakers are far apart in the embedding space. The idea is like word embedding in NLP, where words are represented as fixed-length continuous-valued vectors. Words with similar meanings are close together in the vector space. Again, like word embedding, there are many forms of speaker embeddings. They can be grouped into two categories – unsupervised embedding without the need for labeled data, and supervised embedding that requires labeled data.  An \textit{i}-vector is a form of unsupervised embedding. The \textit{i}-vector model (aka, a total variability model) is a generative model trained using a maximum likelihood criterion. Supervised embeddings rely on discriminative training typically with a multi-class speaker-discriminative loss and the use of labeled data. Recent popular frameworks such as x-vector~\cite{snyder2018x}, r-vector~\cite{zeinali2019but}, and xi-vector~\cite{lee2021xi} belong to this category.

\subsection{Deep Speaker Representation Learning}
In this paper, we will primarily focus on speaker modeling based on deep learning, a technology that has made revolutionary progress in fields such as speech recognition \cite{deng2013new}, image recognition \cite{krizhevsky2012imagenet}, and natural language processing \cite{mikolov2010recurrent}, thanks to advancements in computational power and the powerful nonlinear fitting capabilities of deep neural networks. The widespread application of deep neural networks as deep feature extractors has propelled the field of Representation Learning~\cite{bengio2013representation}, exploring how to utilize networks to learn features that are universal, low-dimensional, and have strong generalization capabilities. The d-vector~\cite{variani2014deep} approach was an early method entirely reliant on neural networks for extracting speaker representations. Despite numerous subsequent enhancements, such as multi-task learning~\cite{chen2015multi}, segmental-level aggregation~\cite{snyder2016deep,li2017deep} and end-to-end metric~\cite{heigold2016end}, the \textit{i}-vector approach remained dominant in speaker modeling until the introduction of x-vectors~\cite{snyder2017deep,snyder2018x}. 
This innovation led to significant improvements in standard speaker recognition datasets. Additionally, the popularity of x-vectors was accelerated by their open-source implementation within the Kaldi~\cite{povey2011kaldi} framework\footnote{While core modules like the temporal pooling layer, which enables segment-level optimization, and softmax-based optimization had been introduced earlier, the open-source recipe in Kaldi significantly advanced the popularity of x-vectors.}. Following this, more researchers explored various neural network architectures~\cite{zeinali2019but,desplanques2020ecapa,wang2023cam++}, training criteria~\cite{cai2018exploring,xiang2019margin,chung2020defence,wang2019discriminative}, and aggregation methods~\cite{cai2018exploring,zhu2018self,wang2021revisiting,xie2019utterance}, leading to the current learning framework characterized by segment-level training and margin-based optimization metrics. Along with the evolution of speaker modeling technology, new applications have emerged, raising new challenges such as robustness, efficiency, utilization of unlabeled data, multi-modality fusion, and more.

\subsection{Unique Perspective of This Overview}

There have been several overview articles on deep learning based speaker recognition~\cite{bai2021speaker,hanifa2021review,kabir2021survey} and speaker diarization tasks~\cite{park2022review}. Among them, the relatively comprehensive one is the paper by Bai et al.~\cite{bai2021speaker}, providing a coverage of mainstream methods for speaker recognition and speaker diarization proposed up until the end of 2020. 

The advent of deep learning marks a paradigm shift in speaker characterization techniques, calling for a new perspective on issues such as learning paradigms, network design, model robustness and interpretability. Furthermore, speaker characterization now has broader applications beyond speaker recognition. This article aims to offer researchers in these fields valuable insights from the lens of speaker modeling.

In this overview, \emph{we do not intend to expend significant effort repeating what has been detailed in previous overview articles}, as they already provide clear explanations. If you wish to learn more about classic technologies before the rise of deep learning, please refer to the paper~\cite{kinnunen2010overview}; if you are interested in the development frameworks for deep learning prior to the year 2020, and the modeling methods related to speaker recognition, we highly recommend the paper~\cite{bai2021speaker}. This overview is different from the prior overview articles in terms of coverage of techniques and the way we discuss and present the techniques. 

\begin{itemize}[leftmargin=2mm]
    \item \textbf{Novel Perspective}: This paper will provide a novel perspective from both theoretical and pratical standpoints, comprehensively reviewing the domain of speaker modeling through the lens of deep speaker representation learning. We will organize the content by different paradigm shifts or research topics in a modularized manner for easier reference.

    \item \textbf{Breadth of Coverage}: This paper will be systematically organized, covering various aspects such as learning paradigms, robustness and interpretation. We would also like to elaborate on how these speaker representations can be applied to various related speech tasks.
    
    \item \textbf{Latest Developments}: This paper will present an up-to-date overview of speaker representation learning that have not been presented in previous overview articles, such as self-supervised learning, large pretrained model, cross-modality learning and speaker interpretable learning.
\end{itemize}

\vspace{-1em}

\subsection{Organization of This Overview}


\tikzstyle{leaf}=[mybox,minimum height=1.2em,
fill=hidden-orange!50, text width=5em,  text=black,align=left,font=\footnotesize,
inner xsep=4pt,
inner ysep=1pt,
]

\begin{figure*}[thp]
  \centering
  \begin{forest}
    forked edges,
    for tree={
      grow=east,
      reversed=true,  
      anchor=base west,
      parent anchor=east,
      child anchor=west,
      base=left,
      font=\normalsize,
      rectangle,
      draw=hiddendraw,
      rounded corners,
      align=left,
      minimum width=2.5em,
      inner xsep=4pt,
      inner ysep=0pt,
    },
    where level=1{text width=14.5em,align=center,font=\normalsize}{},
    where level=2{text width=19.5em,align=left,font=\normalsize}{},
    [Speaker Modeling
      [Fundamentals
        [Sec.~\ref{sec:intro}: Introduction]
        [Sec.~\ref{sec:formulation}: Terminology and Task Formulation]
        [Sec.~\ref{sec:encoder}: Speaker Encoders]
      ]
      [Sec.~\ref{sec:paradigm}: Paradigm Shift
        [Sec.~\ref{sec:ssl}: Supervised to Self-supervised]
        [Sec.~\ref{ssec:semi_supervised_training}: Semi-superivised Training]
        [Sec.~\ref{sec:pretrain}: Leverage Pretrained Models]
        [Sec.~\ref{sec:modality}: Unimodal to Multi-modal]
      ]
      [Sec.~\ref{sec:robustness}: Robustness \& Efficiency
        [Sec.~\ref{sec:text}: Text Dependency]
        [Sec.~\ref{sec:language}: Cross-language Recognition]
        [Sec.~\ref{sec:acoustic}: Channel Variability]
        [Sec.~\ref{sec:speaker}: Intra-speaker Variability]
        [Sec.~\ref{sec:duration}: Duration Variability]
        [Sec.~\ref{sec:noisy_label}: Learning from Noisy Labels]
        [Sec.~\ref{sec:quantization}: Model Quantization]
        [Sec.~\ref{sec:kd}:
        Knowledege Distillation]
        [Sec.~\ref{sec:effarch}: Efficient Architectures]
        [Sec.~\ref{sec:effpolicy}: Efficient Policy]
      ]
      [Sec.~\ref{sec:explain}: Explainablity
        [Sec.~\ref{sec:probe}: Probing Tasks]
        [Sec.~\ref{sec:ablation}: Component Ablation]
        [Sec.~\ref{sec:relevance}: Relevance analysis]
        [Sec.~\ref{sec:visualize}: Visualization]
        [Sec.~\ref{sec:xai}: Explainable Network Design]
      ]
      [Sec.~\ref{sec:application}: Applications \& Strategies
        [Sec.~\ref{sec:integration}: Integration Strategies]
        [Sec.~\ref{sec:diarization}: Speaker Diarization]
        [Sec.~\ref{sec:asr}: Speech Recognition]
        [Sec.~\ref{sec:generation}: Speech Generation]
        [Sec.~\ref{sec:tse}: Target Speaker Front-Ends]
      ]
      [Sec.~\ref{sec:open}: Dataset and Tools
        [Sec.~\ref{sec:data}: Dataset]
        [Sec.~\ref{sec:tool}: Open-Source Toolkits]
      ]
      [Sec.~\ref{sec:trends}: Trends
        [Sec.~\ref{sec:privacy}: Privacy and Ethics]
        [Sec.~\ref{sec:lssl}: Large-Scale SSL]
        [Sec.~\ref{sec:cross}: Speaker-Text Cross Modality]
        [Sec.~\ref{sec:con_attack}: Attacks and Defenses]
        [Sec.~\ref{sec:con_spkgen}: Controlable Speaker Generation]
        [Sec.~\ref{sec:disentangle}: Disentanglement Learning]
      ]
      [Sec.~\ref{sec:conclusion}: Conclusion]
    ]
  \end{forest}

\caption{Organization of this paper}
\label{org_survey_paper}
\end{figure*}
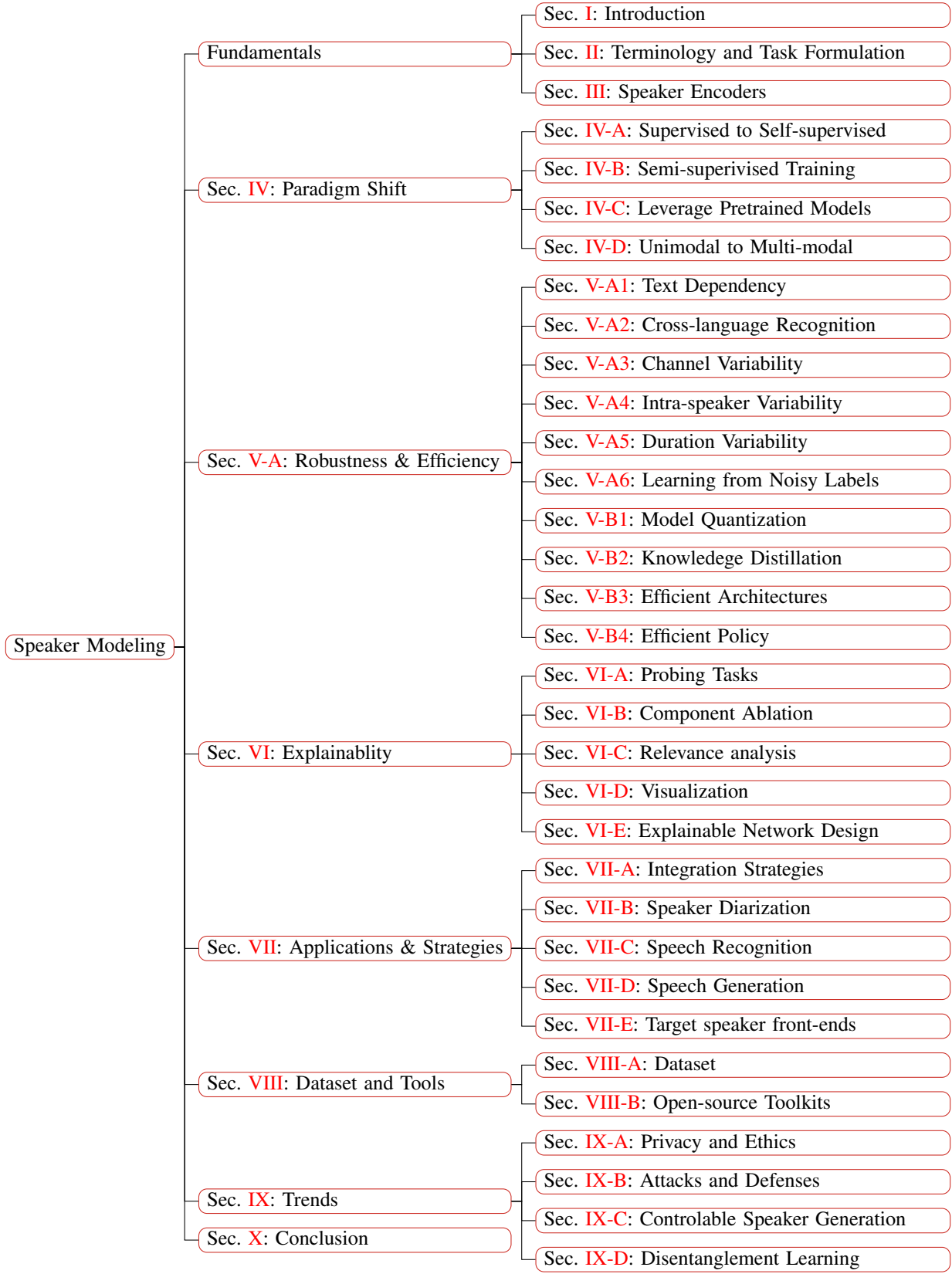

This review paper is organized as in Figure~\ref{org_survey_paper}. Firstly, we will introduce fundamental concepts and classical techniques, which are primarily covered in Sections~\ref{sec:intro}, \ref{sec:formulation}, and \ref{sec:encoder}. We will then explore the field from various perspectives, starting from the recent paradigm shift in deep speaker representation learning. Section~\ref{sec:paradigm} encompasses changes in learning paradigms such as the growing prominence of self-supervised learning, the utilization of powerful pre-training models, multimodal fusion, and more. Next, in Section~\ref{sec:robust_eff}, we will delve into robustness and efficiency issues, which are critical for real-world applications.


Moreover, similar to the rise of Explainable AI (XAI) research in other domains, the field of speaker modeling has witnessed the emergence of similar work. Section~\ref{sec:explain} will be dedicated to describing the efforts toward explainable speaker modeling and exploring the underlying mechanisms. 

As previously mentioned, speaker modeling extends beyond traditional tasks like speaker recognition and diarization, finding application in various scenarios that require speaker modeling. In Section~\ref{sec:application}, we will provide a detailed overview of relevant applications. Rather than simply introducing these applications, we will explain how speaker modeling is integrated into them, including the utilization of pre-trained speaker models and customized joint training models.

Next, in Section~\ref{sec:open}, we will briefly introduce common datasets and open-source tools for the convenience of researchers seeking quick references, as they have significantly advanced the development of speaker related technologies.

In Section~\ref{sec:trends}, we will highlight several research trends that are important but not yet fully explored. Finally, Section~\ref{sec:conclusion} concludes this overview article.

Overall, through this modularized design, our aim is to enable readers to easily navigate to the sections they find most relevant. Additionally, we seek to provide researchers who are not directly involved in the field of speaker modeling with a clearer understanding of how speaker modeling can enhance their own related research endeavors.
\vspace{-1em}

\section{Formulation of The Deep Speaker Representation Learning Problem}
\label{sec:formulation}
Speaker representation learning is the process of extracting compact and discriminative representations from given speech signals. These representations aim to capture the unique acoustic characteristics of speakers and maintain consistency across different languages, content, and environmental conditions. Mathematically, this can be defined as a mapping function $f$ that maps a speech signal  $X = \{x_1, x_2, \ldots, x_T\} \in \mathbb{R}^T$  of length $T$ to a fixed-length vector $\mathbf{v}$, which represents the speaker's identity:
\begin{equation}
\mathbf{v} = f(X; \Theta)
\end{equation}

\vspace{-0.5em}
To standardize terminology and avoid ambiguity in the context of deep speaker representation learning, we define $f$ as the \emph{speaker encoder}, which is based on a deep neural network and parameterized by $\Theta$. The learned representation $\mathbf{v}$ is often referred to as the \emph{speaker embedding} due to its vector format.

Ideally, $\mathbf{v}$ learnt from $f$ should satisfy:
\begin{equation}
\text{sim}(\mathbf{v}_{i,m}, \mathbf{v}_{i,n}) > \text{sim}(\mathbf{v}_{i,m}, \mathbf{v}_{j,k}) \quad \forall i \neq j, \forall m, n, k
\end{equation}
That is, the similarity between any two representations of utterances from speaker $i$ should be higher than the similarity between any utterance from speaker $i$ and any utterance from speaker $j$. The most common definition of the similarity function $\textbf{sim}$ is cosine similarity: $\text{sim}(\mathbf{a}, \mathbf{b}) = \frac{\mathbf{a} \cdot \mathbf{b}}{|\mathbf{a}| |\mathbf{b}|}$.

However, achieving the above conditions in practical representation learning is challenging. We can further decompose the learning objectives as follows:
\begin{enumerate}
\item \textbf{Discrimination}: The representation vectors of different speakers should be as dissimilar as possible, reflecting a large inter-class variance.
\item \textbf{Consistency (Robustness)}: For different utterances from the same speaker, even under different speech contents, environmental noise, and channel conditions, their representation vectors should remain consistent, reflecting a small intra-class variance.
\item \textbf{Compactness}: The dimensionality of the representation vectors should be as small as possible for storage and computational efficiency, while containing sufficient information for effective identification.
\end{enumerate}
\section{Speaker Encoders}
\label{sec:encoder}
We will start by reviewing  various neural architectures to learn speaker representations~\cite{snyder2018x, zeinali2019but, gulati2020conformer, zhang2022mfa}, and then will discuss the prevalent encoder architectures that are used in different learning paradigms. 


Speech signals are typically processed by framing, which creates a discrepancy between frame-level input features and the desired segment-level speaker representation. Consequently, speaker embedding learning methods can be categorized based on whether optimization occurs at the frame level or the segment level. Frame-level methods treat aggregation as a post-processing step, while segment-level methods integrate aggregation within the neural network for direct optimization.

\subsection{Speaker Encoders Optimized at Frame-Level}

\begin{figure}[ht!]
  \centering
  \includegraphics[width=0.4\textwidth]{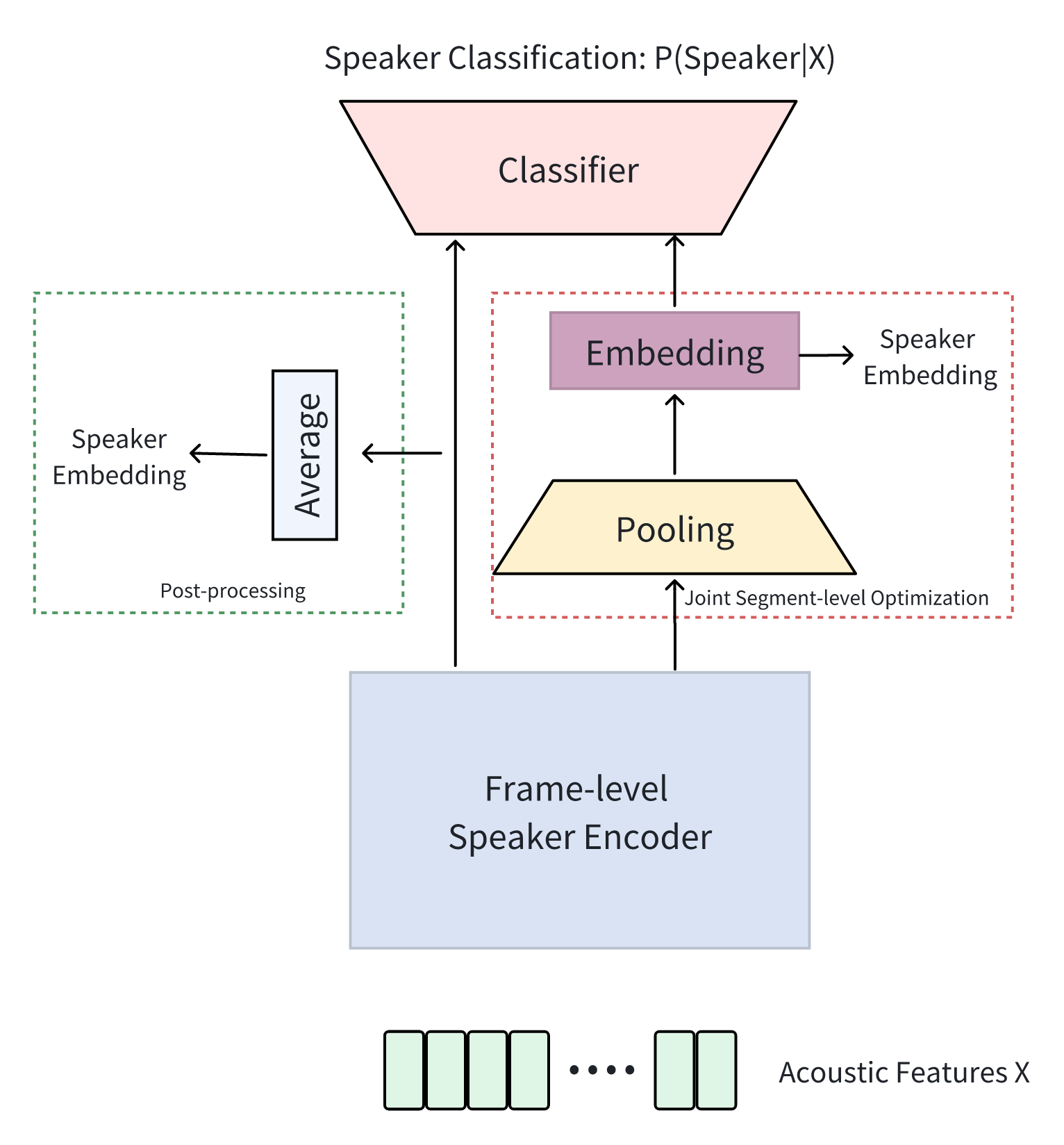}
  \caption{Frame-level optimization v.s. Segment-level optimization in a typical discriminative speaker embedding learning framework}
  \label{fig:discri}
\end{figure}

Represented by d-vector~\cite{variani2014deep}, the process of speaker representation learning optimized at the frame level is illustrated in Figure \ref{fig:discri}. The neural network takes frame-level features as input and optimizes the cross-entropy (CE) loss function at this granularity. Once the network training is complete, for an input speech feature comprising $T$ frames, denoted by $\mathbf{O} = \{\mathbf{o}_1, \cdots, \mathbf{o}_T\} \in \mathbb{R}^ {T \times D}$, we extract frame-level outputs from the hidden layers close to the output layer, represented by $\mathbf{F} = \{\mathbf{f}_1, \cdots, \mathbf{f}_T\} \in \mathbb{R}^ {T \times D'}$. Then, an aggregation operation is applied to transform the sequence of frame-level representations into a sentence-level speaker representation. The most common method of aggregation is averaging:

\begin{equation}
\mathbf{v}_{\mathrm{dvec}} = \frac{1}{T}\sum_{t=1}^T \mathbf{f}_t
\end{equation}

The d-vector was a notable attempt based on NN; however, its simplistic network structure and frame-level optimization criteria prevented it from large-scale adoption.

\subsection{Speaker Encoders Optimized at Segment-Level}

The problem of sample granularity mismatch during the training and usage phases, as observed with the d-vector, can be addressed through segment-level optimization methods: during training, explicitly binding the features of different frames from the same utterance and optimizing directly at the entire speech segment level. Typically, the method of binding different speech frames involves introducing an aggregation layer into the neural network that maps the sequence of frame-level features to a segment-level vector representation. 
\subsubsection{Aggregation Layers}
For speaker representation, common aggregation layers include relatively simple statistical-based methods and more complex methods that employ operations such as attention mechanisms \cite{bahdanau2014neural, liu2018exploring, zhu2018self, wang2018attention}, and dictionary learning\cite{cai2018exploring}. Temporal Average Pooling (TAP) and Temporal Statistics Pooling (TSTP) are the two most common aggregation methods in deep speaker representation learning. TAP calculates the mean of the deep features of the frame sequence to represent the segment level, while TSTP also considers variance information by concatenating the mean and standard deviation vectors to form the segment-level representation.
Generally, TSTP performs better than TAP in speaker verification tasks. ~\cite{wang2018covariance} and ~\cite{wang2021revisiting} shows the effectiveness of the pure variance, leading to Temporal Standard Deviation Pooling (TSDP). Higher-order statistics were also discussed in ~\cite{wang2021revisiting}, but no performance improvement was observed.

\subsubsection{Time-Delay Neural Networks}
x-vector~\cite{snyder2017deep,snyder2018x} is the most popular form of speaker representation based on segment-level optimization methods. Compared to the d-vector, it not only introduces segment-level optimization but also utilizes a Time Delay Neural Network (TDNN) with greater modeling capacity as the speaker representation extractor. This structure dates back to its application in phoneme recognition tasks in 1989\cite{waibel1989phoneme}. The TDNN is based on one-dimensional convolution (1-D Convolution), manifesting as a feedforward neural network. It can be conceptualized as a collection of hierarchical subnetworks, with each subnetwork progressively expanding its receptive field. The x-vector, as a representative of segment-level speaker representation frameworks, was the first widely adopted DNN based speaker embedding.
ECAPA-TDNN (Emphasized Channel Attention, Propagation and Aggregation in TDNN)~\cite{desplanques2020ecapa} is an improvement over the traditional TDNN, and the main improvements can be summarized as follows: 1) Channel attention mechanism which automatically highlights important speaker-specific information; 2) Multi-scale feature learning to capture feature patterns at different time resolutions; 3) Multi-level feature aggregation to leverage information in different layers; 4) Residual connections to facilitate better gradient flow; 5) Dense connections to better preserve information from earlier layers. These improvements to TDNN as seen in ECAPA-TDNN actually reflect the main directions in which speaker encoders have evolved in recent years. We will discuss this in more detail in Section~\ref{sec:tax_speaker_encoder}.

\subsubsection{Residual Networks}

However, researchers have come to the realization that indiscriminately increasing model depth does not guarantee performance enhancement. The advent of Deep Residual Neural Networks (ResNet), as introduced by He et al.~\cite{he2016deep}, stands as a pivotal advancement in the realm of image recognition and the broader field of deep learning. ResNet addresses the issues of vanishing or exploding gradients encountered during the training of deep neural networks through the introduction of residual learning. This allows for the construction of deeper network architectures, thus enhancing model performance. Since its introduction, ResNet has significantly improved the state-of-the-art across numerous image-related tasks and has gradually been adopted by researchers for speaker modeling tasks. For example, in~\cite{chung18b_interspeech}, the baseline systems provided by the authors were based on ResNet34 and ResNet50, but largely adopted the architecture used in image processing, resulting in mediocre performance. Li et al.~\cite{lina2018deep} introduced the Inception-ResNet structure and explored its robustness under different durations. In the work~\cite{cai2018exploring} and ~\cite{zeinali2019but}, although the ResNet structures used varied in parameter settings, all removed the shallow pooling modules of ResNet from previous studies. This configuration has now become widely used and is referred to as r-vector in \cite{zeinali2019but}. The r-vector won the championship in the 2019 VoxSRC competition and further promoted the widespread adoption of this architecture. Subsequently, various variants of ResNet, such as Res2Net~\cite{zhou2021resnext,roy2022res2net}, ResNext~\cite{zhou2021resnext,yan2024gmm}, DF-ResNet~\cite{liu2022df,liu2024golden}, ERes2Net~\cite{chen2023enhanced}, have been explored under the speaker modeling framework. In the recent VoxSRC 2023 competition\footnote{\url{https://mm.kaist.ac.kr/datasets/voxceleb/voxsrc/competition2023.html}}, the winning solution even expanded this structure to over 500 layers~\cite{zheng2023unisound}.

\subsubsection{Transformer Based Models}
Transformer was proposed in~\cite{Vaswani2017attention} and it has recently demonstrated exemplary performance on a broad range of natural language processing (NLP)~\cite{devlin2018bert}, compute vision (CV)~\cite{dosovitskiy2020image}, and speech-related~\cite{gulati2020conformer} tasks. Compared with tradition backbones based on recurrent neural networks (RNNs) and convolutional neural networks (CNNs), the advantage of self-attention in Transformer lies in its powerful global information modeling capability and parallel computation ability~\cite{Vaswani2017attention}. 
In order to explore the modeling ability of transformer for speaker features, researchers in~\cite{mary2021s, safari2020self} attempted to introduce vanilla transformer into speaker verification tasks for the first time. However, due to the lack of transformer's ability to model local features, the performance was not satisfactory. To alleviate this problem, local attention mechanism by restricting the receptive field of the attention heads and incorporating with CNNs are exploring in~\cite{han2022local,wang2022multi} to capture global dependencies and model the locality. Subsequently, there was a lot of work exploring how to better integrate transformer and CNN, each with their own strengths. Some of these works attempt to insert CNNs into self-attention modules~\cite{zhang2022mfa,choi2022improved,wang2023lightweight,sang2023improving}, while others use dual branch modeling and then cross fusion two branches~\cite{yao2023branch,wang2023p,sun2024branchformer}. At this point, the transformer based models have achieved comparable performance to the convolutional models.

\subsection{Frame-Level v.s Segment-Level}

NN-based speaker encoders have shifted from frame-level to utterance-level in recent years, such segment-level model architecture and optimization granularity are more suitable for most application scenarios. However, there are still some scenarios where there is a need for modeling frame-level speaker representations. For example, the paper~\cite{tawara2020frame} introduced the segment-level multi-task and adversarial training methods from the previous work~\cite{wang2019usage} into frame-level speaker encoders and demonstrated that frame-level speaker representations can be more effective in short-term speaker recognition tasks (such as those less than 1.4 seconds). Intuitively, frame-level fine-grained embeddings should also be more conducive to short-duration scenarios. Similarly, in speaker diarization tasks, where such as EEND~\cite{maiti2023eend} and TS-VAD~\cite{medennikov2020target}, frame-level speaker modeling is involved. This is because speaker diarization requires very precise timestamped prediction of speaker identity.

\subsection{1D Convolution v.s. 2D Convolution}
Speech is naturally one-dimensional sequential signal. Therefore, for model architectures that use raw waveforms as input, such as RawNet~\cite{jung2019rawnet,jung2020improved,jung2022pushing} and SincNet~\cite{ravanelli2018speaker}, one-dimensional convolution-based backbones are often employed. However, for time-frequency features, such as spectrograms or filter banks (Fbank), the model input is transformed into a two-dimensional format similar to images. In this case, both one-dimensional and two-dimensional convolutions can be applied.

Models represented by the TDNN series~\cite{snyder2017deep,snyder2018x,desplanques2020ecapa} use one-dimensional convolutions, typically applied along the time dimension. One-dimensional convolutions have lower computational requirements, making them suitable for quickly processing long time series data. They can also use relatively large convolutional kernels (such as dilated convolutions in TDNN) to capture long-term dependencies, and the overall model structure is relatively simple. However, there are also some drawbacks: only convolving along the time dimension may not fully capture the interrelationships between different frequencies, nor can it simultaneously capture complex dependencies in both time and frequency dimensions.

Two-dimensional convolutions are applied along both the time and the frequency dimension, which can compensate for the aforementioned shortcomings of one-dimensional convolutions. This helps in better modeling the time-frequency structure of speech signals and often enhances the model's capacity. Correspondingly, such models, represented by ResNet~\cite{cai2018exploring,zeinali2019but,zhou2021resnext}, require more computational resources and memory, and the training time is relatively longer. Additionally, the higher complexity of these models may lead to overfitting problems, necessitating a larger amount of training data and having a higher performance ceiling.

\begin{figure}[ht!]
  \centering
  \includegraphics[width=0.25\textwidth]{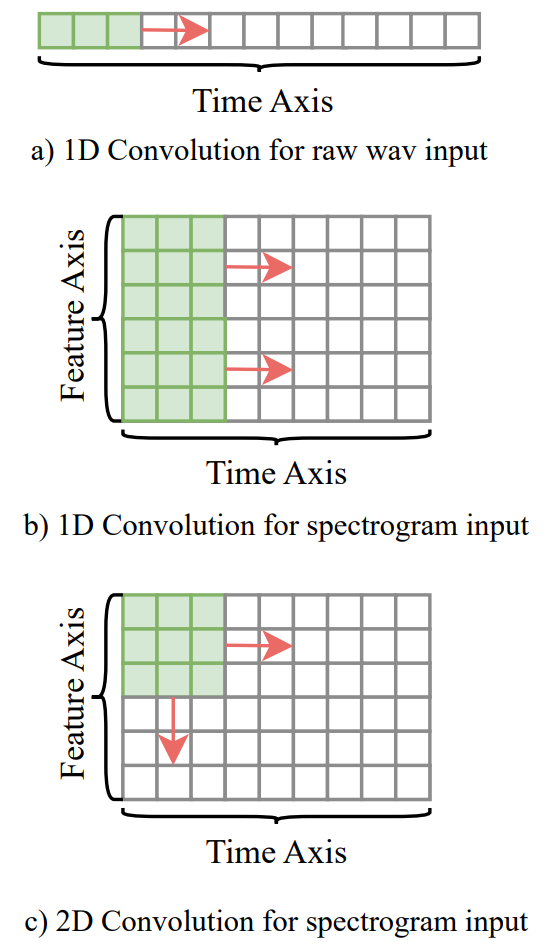}
  \caption{1D v.s. 2D convolution in Speaker Encoders}
  \label{fig:convoluation}
\end{figure}

Therefore, the choice of which convolution method to use depends on the specific task requirements, computational resources, and the size of the dataset. However, from experience, using two-dimensional convolutions in layers directly related to two-dimensional input spectrograms might potentially improve performance. For example, one of the significant improvements in the latest ECAPA-TDNN2~\cite{thienpondt2023ecapa2} is the introduction of 2D convolutions in the shallow layers, similar approach can also be found in ECAPA-CNN-TDNN~\cite{thienpondt2021integrating}, MFA-TDNN~\cite{liu2022mfa}, where a 2D-CNN-based module is appended in front of the original ECAPA-TDNN.

\subsection{Taxonomies of Speaker Encoder Improvement Approaches}
\label{sec:tax_speaker_encoder}
In the following, we summarize the recent directions of architectural improvements.
\subsubsection{Deepening Network Architecture}
Commonly, various highway connections, such as residual connections~\cite{zeinali2019but,desplanques2020ecapa} and dense connections~\cite{yu2020densely}, are employed here to mitigate the issue of vanishing gradients, thereby facilitating the construction of deeper neural network architectures.

\subsubsection{Utilizing Contextual Information}
Initially for speaker modeling, contextual information typically refers to the size of the input window that the model can perceive. For instance, models based on pure DNN structures like d-vectors often employ frame-extension to enlarge the receptive field of the current frame being modeled~\cite{variani2014deep}. TDNN~\cite{snyder2017deep} addresses this through dilated CNNs, which provide a more flexible approach. However, the classic x-vector~\cite{snyder2018x} setup has a relatively limited receptive field, covering at most tens of frames of contextual information, and thus is still considered local information. More recent architectures, such as transformers, are inherently capable of modeling long-distance dependencies due to the self-attention mechnism. To facilitate more efficient utilization of contextual information, the following three methods are commonly adopted:
\begin{itemize}
    \item \emph{Multi-scale modeling}: Different subnets operate at different scales or on different feature bins. These subnets can process either the direct input \cite{zhu2020vector} or intermediate features \cite{liu2022mfa,mun2023frequency,desplanques2020ecapa,roy2022res2net,zhou2021resnext,heo2024next,liu2024rep,li2021si}. 
    \item \emph{Cross-layer aggregation}: The scale and time resolution across different layers are usually distinct. Integrating features hierarchically~\cite{tang2019deep,jung2020improving,desplanques2020ecapa,zhang2022mfa} can also enhance the utilization of contextual features.
    \item \emph{Explict local and global information modeling}:  In certain scenarios, it is beneficial to utilize longer speech signals. Consequently, some efforts have been made to incorporate more global contextual information into modeling \cite{xie2022global,han2022local,liu2022mfa,han2022mlp,zi2024resformer,li2024ds}. This approach has the potential to improve fault tolerance to noise and impaired audio in long-duration speech. 
\end{itemize}

\subsubsection{Automatic Feature Selection and Reweighting}
Features in different frequency bins contain varying information \cite{lu2008investigation}, and their importance is not uniform for speaker modeling. Attentive pooling can be regarded as automatic reweighting along the time axis, while similar reweighting and automatic feature selection techniques can be found along the frequency or channel axis~\cite{deng2022importance,gu2023dynamic,shi2020robust,kataria2020feature,hajavi2020knowing}.

\section{Shift of Learning Paradigms}
\label{sec:paradigm}

In Section~\ref{sec:intro}, we summarized the long-term paradigm shifts that have occurred over the past two decades. In this section, we will focus on the paradigm shifts that have taken place in the last several years.
While previous models were primarily trained from scratch in a supervised manner, this section explores recent advancements in self-supervised learning~\cite{chen2020simple, MOCO, dino, heo2022self, cho2022non, zhang2022c3, cai2021iterative, han2022self, chen2023comprehensive} and pretrained large-scale speech models~\cite{Baevski2019vqwav2vecSL, baevski2020wav2vec, hsu2021hubert, chen2022wavlm, chen2022large}. Additionally, we will discuss progress in multi-modality and cross-modality learning.

\begin{figure*}[ht!]
  \centering
  \includegraphics[width=0.95\textwidth]{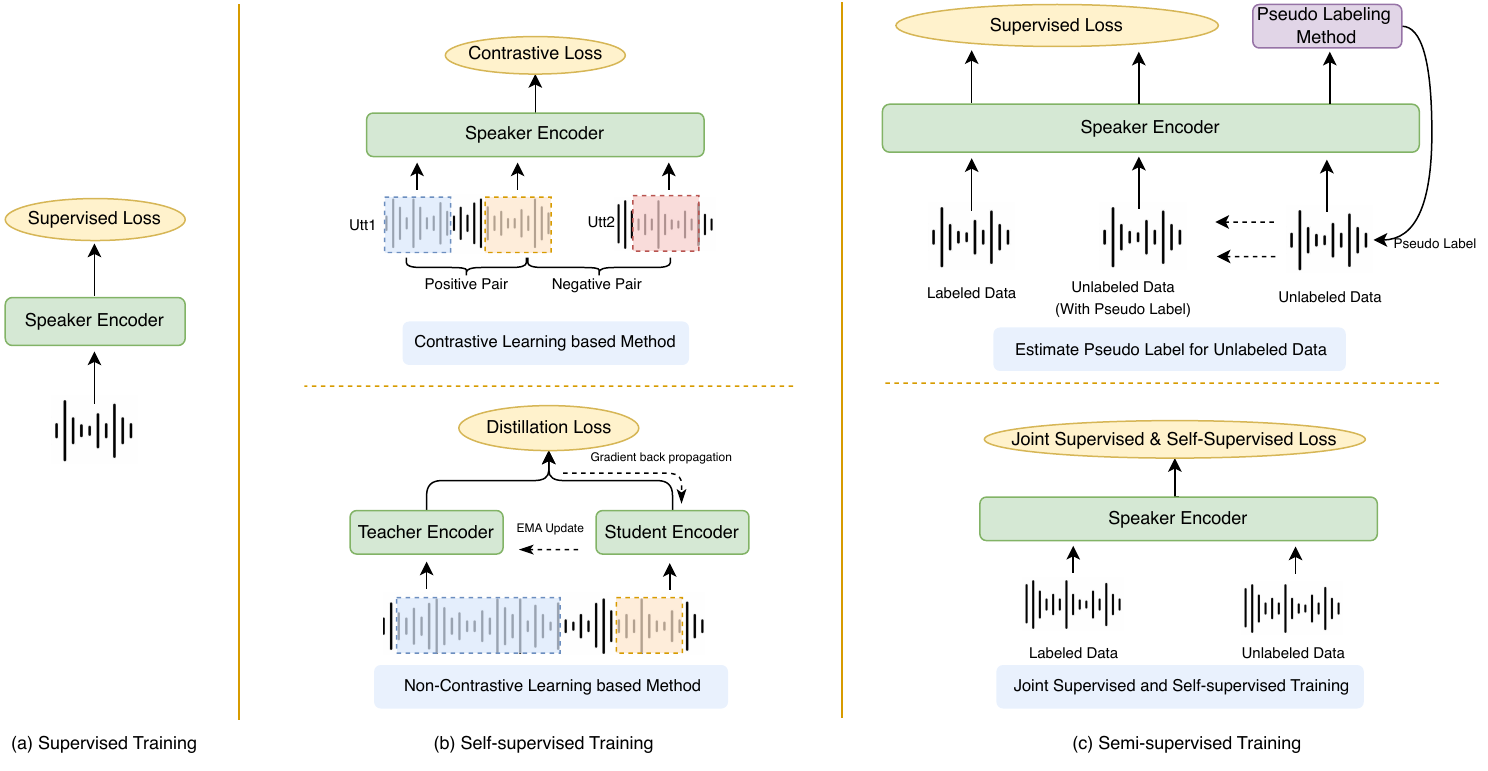}
  \caption{Shift of Learning Paradigms}
  \label{fig:supervised_self_semi}
\end{figure*}

\subsection{From Supervised To Self-Supervised Learning}
\label{sec:ssl}

\subsubsection{Supervised Learning Methods}

In supervised learning, training data must include both inputs and expected outputs (labels). When the dataset is sufficiently large and accurately labeled, this method can often achieve relatively precise modeling results. Within the context of speaker representation learning, the loss functions for supervised training can be broadly classified into two categories: loss functions based on softmax classification and end-to-end loss functions based on metric learning.

\noindent \textbf{Classification Based Objectives:} Softmax loss, a commonly utilized classification loss function for training speaker-discriminative deep neural networks (DNNs), can be formulated as:
$$L_\text{softmax}=-\frac{1}{N}\sum_{i=1}^N\log \frac{e^{\mathbf{W}^T_{y_i}\mathbf{x}_i+\mathbf{b}_{y_i}}}{\sum_{j=1}^c e^{\mathbf{W}^T_{j}\mathbf{x}_i+\mathbf{b}_{j}}}$$
where $N$ is the batch size, $c$ is the number of classes. $\mathbf{x}_i\in \mathbb{R}^d$ denotes the $i$-th input of samples to the projection layer and $y_i$ is the corresponding label index.
$\mathbf{W}\in \mathbb{R}^{d\times c}$ and $\mathbf{b}\in \mathbb{R}^{c}$ are the weight matrix and bias in the projection layer.
Although Softmax loss effectively penalizes classification errors, it does not explicitly optimize for within-class compactness or between-class separability. This lack of explicit regulation can result in suboptimal performance in speaker recognition tasks, thereby motivating researchers to develop optimization objectives that impose explicit constraints on these aspects. Enhanced variants, such as A-Softmax~\cite{li2018angular, huang2018angular}, AM-Softmax~\cite{wang2018additive, yu2019ensemble}, and AAM-Softmax~\cite{deng2019arcface, xiang2019margin}, address inter-class distances by introducing margins in angular space or cosine similarity, consequently improving inter-class discrimination. Detailed comparison of these margin-based Softmax for speaker embedding learning can be found in ~\cite{xiang2019margin}.

\noindent \textbf{Metric Learning Based Objectives:}
Besides the classification based training objective, metric learning-based loss functions have also been investigated for the task of speaker representation learning. Triplet Loss~\cite{schroff2015triplet,zhang2017end,zhang2018text,huang2018joint} and Quadruplet Loss~\cite{chen2017beyond,narayanaswamy2019designing} focus on the relative distances between positive sample pairs and negative pairs, while Center Loss~\cite{wen2016discriminative,cai2018exploring,wang2019discriminative} emphasizes minimizing intra-class variations and is typically used in combination with traditional softmax loss to achieve a balance between inter-class separability and intra-class compactness.

\subsubsection{Self-Supervised Methods}
\label{sssec:self_supervised_method}
Collecting large-scale dataset with speaker labels is time-consuming and may pose privacy issues sometimes. Thus, discovering potential labels and internal structure from the data itself and designing effective self-supervised training methods become more and more necessary. The self-supervised learning method can be roughly divided into two categories: generative and discriminative. The generative approach enables the model to learn specific information through reconstructing the input data \cite{hinton2006reducing,pathak2016context}. In \cite{stafylakis19_interspeech}, the proposed system can learn speaker embedding without speaker labels by reconstructing the input audio waveform. In self-supervised speaker representation learning field, most reseachers focus on the discriminative methods:

\noindent \textbf{Contrastive Learning Based Method:}
Contrastive learning in self-supervised setups is akin to metric-based methods, but with two basic assumptions: 1) There is only one speaker in one utterance or short consecutive interval 2) Different utterances contain different speakers. Based on the first assumption, researchers can sample two segments from the same speaker, which forms a positive pair, $(x_i, x_i^{+})$. Based on the second assumption, the negative pairs, $(x_i, x_i^{-})$, can be found when two segments are from different utterances. Then, researchers can design effective contrastive loss functions to enlarge the distance between the negative pairs and minimize the distance between positive pairs:

\begin{equation}
\mathcal{L_{\textit{con}}}=\sum_{N^{+}} \operatorname{d}\left(x_i, x_i^{+}\right) - \sum_{N^{-}} \operatorname{d}\left(x_i, x_i^{-}\right)
\end{equation}
where $\operatorname{d}(\cdot, \cdot)$ can be any distance metric formula, and the loss function does not have to be as simple as the one above. However, the ultimate goal is to make the distance between positive pairs smaller and the distance between negative pairs larger. For instance, Jati et al. \cite{jati2019neural} first measured the distance between two segments using $L_1$ distance and then used a binary classification loss to distinguish the positive and negative pair. Ravanelli et al. \cite{ravanelli19_interspeech} maximize the mutual information (MI) between positive pairs and minimize it between negative pairs. To make the model more robust to the channel variation, the authors \cite{zhang2021contrastive} added different data augmentations to the sampled segments from the same utterance and added an extra loss function to constrain the distance for the positive pairs. Besides, the authors in \cite{huh2020augmentation} leveraged extra adversarial training loss to help the system more robust to the noise. To enrich the diversity of the negative pair data in one batch, Xia et al. \cite{xia2021self} employed a buffer to store the speaker embedding samples from the previous batches.

\noindent \textbf{Non-Contrastive Learning Based Method:}
Although the contrastive learning based methods have shown effectiveness in learning speaker representation from unlabeled data, the assumption, ``different utterances contain different speakers'', may bring false negative pairs, in which two segments from different utterances are from the same speaker. And the authors in \cite{han_cluster-dino_2024} pointed out that almost every batch contains at least one false negative pair when the batch size becomes 256 for the Voxceleb2 dataset \cite{nagrani2020voxceleb}.
To solve this problem, `self-\textbf{di}stillation with \textbf{no} labels (DINO)'~\cite{caron2021emerging,jung22_interspeech,heo2022self,chen2023comprehensive} strategy is proposed. In the DINO strategy, as illustrated in the middle bottom part of Figure \ref{fig:supervised_self_semi}, there are two parallel networks, the student network and the teacher network. The two segments from the same utterance, i.e. positive pair, are fed into the student and teacher network respectively. The student network and teacher network map the input into a high-dimensional distribution, and then the loss is defined to minimize the divergence between two output distirbutions:
\begin{equation}
\mathcal{L}_{\textit{DINO}} = \operatorname{CrossEntropy}\left(\textit{student}(x_i), \textit{teacher}(x_i^+)\right)
\end{equation}
In the system optimization process, the student network is updated by the gradient backpropagation and the teacher network is moving averaged from the student network.
Jung et al. leveraged the `self-\textbf{di}stillation with \textbf{no} labels (DINO)' \cite{caron2021emerging} in a raw waveform based system and outperformed the previous self-supervised learning methods \cite{jung22_interspeech}. 
Besides, the authors in \cite{heo2022self} found that, based on the DINO strategy, gradually adding more speakers for the training can further improve performance. Further, Chen et al. \cite{chen2023comprehensive} gave a comprehensive analysis of the DINO-based method of speaker representation learning task by analyzing the effect of data augmentation, speaker diversity, and number of the sampled segments.
Despite that the above methods have shown that the non-contrastive learning methods outperformed the contrastive learning methods a lot, Zhang et al. \cite{zhang2022c3} found that combining both methods can further improve the performance.
The DINO-based non-contrastive learning method aims to learn some consistent information within one utterance. The researches described above only focus on applying DINO to speaker representation learning related exploration, and the authors in \cite{cho2022non} found that such a method is also useful in emotion recognition and Alzheimer’s disease detection.

\noindent \textbf{Iterative Model Refinement:}
Although methods based on non-contrastive learning can greatly enhance the performance of the model, they still can't make the model reach a comparable performance as the fully-supervised systems. The performance gap between the supervised training methods and self-supervised learning methods is still large. To mitigate this performance gap, researchers have proposed the iterative refinement strategy \cite{thienpondt2020idlab,cho2021jhu}. In such a strategy, the self-supervised pre-trained model is considered a seed model to extract speaker embedding for each utterance, and then specific clustering methods are applied to assign a unique pseudo-speaker label for each utterance. Based on the pseudo-speaker label, supervised training methods are applied to train a better model. Next, embedding clustering is applied again based on this better model to refine the pseudo label further. We can iteratively do this process until the model performance converges. In this iterative refinement process, the quality of the pseudo label determines the performance of the model. Apart from getting the pseudo label from the self-supervised speaker representation learning model, Chen et al. \cite{chen2023unsupervised} also tried to get the pseudo label from the general speech pre-trained model, and  Cai et al. \cite{cai2022incorporating} leverage extra visual information to extract better pseudo labels. Besides, Tao et al. \cite{tao2022self} and Han et al. \cite{han2022self} have proposed loss-gate strategies to detect unreliable pseudo labels in the supervised training process and achieved further performance improvement.

\subsection{Semi-Superivised Training}
\label{ssec:semi_supervised_training}
In real applications, there is typically a small amount of labeled data and a large quantity of unlabeled data, characterizing a semi-supervised scenario. The most frequently adopted methods~\cite{zheng19c_interspeech,qin20222022,li2024multi} involve initially using the labeled data to pre-train a model. This model is then employed to generate pseudo-labels for the unlabeled data. Subsequently, both the labeled data and the unlabeled data with pseudo-labels are combined to train a new model. However, the accuracy of the pseudo-labels plays a crucial role in the final performance, and the quantity of labeled data determines the effectiveness of the pre-trained model in generating pseudo-labels. Furthermore, Inoue et al.\cite{inoue2020semi} and Choi et al.\cite{choi2023extending} proposed combining the self-supervised and supervised training objectives into a single generalized objective. This approach allows the system to be jointly trained with both labeled and unlabeled data. Typically, the speaker recognition system involves two stages, the speaker embedding extractor training stage and the system inference stage. Semi-supervised methods are typically applied during the speaker embedding extractor training stage. Chen et al.~\cite{chen21v_interspeech} introduced a graph-based label propagation method that leverages both labeled enrollment data and additional unlabeled data during the inference stage to enhance speaker recognition performance in household smart speaker scenarios.

\begin{figure*}[ht!]
  \centering
  \includegraphics[width=0.75\textwidth]{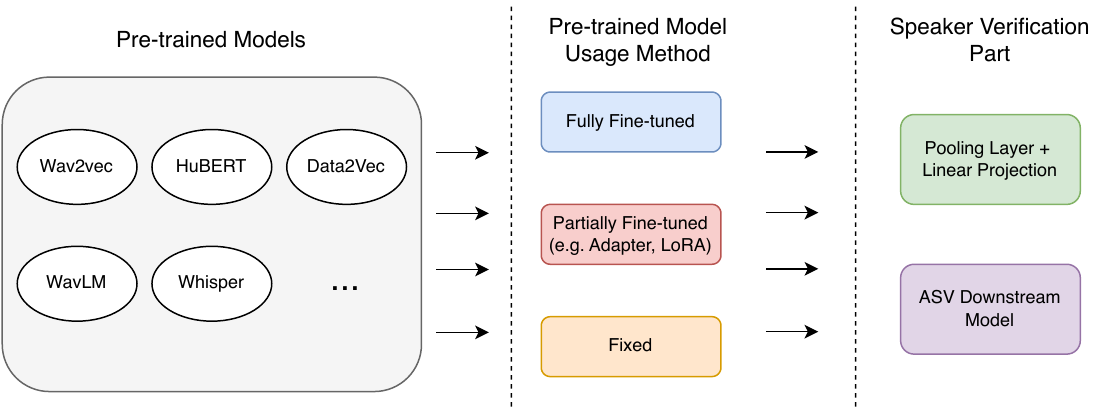}
  \caption{Leveraging Large Pre-trained Models in Speaker Verification task}
  \label{fig:leverage_pretrain}
\end{figure*}

\subsection{Leveraging Large Pre-trained Models}
\label{sec:pretrain}
In recent years, large-scale self-supervised speech pre-trained models \cite{schneider2019wav2vec,baevski2020wav2vec,hsu2021hubert,chen2022wavlm,baevski2022data2vec} have gained a lot of attention, where the researchers first pre-trained a model on the large-scale unlabeled dataset and then applied the model to different speech-related downstream tasks. The researchers in \cite{yang21c_interspeech} elaborately designed a benchmark, called SUPERB, to evaluate the pre-trained model on different downstream tasks. Because the paradigms of different downstream tasks are different, the corresponding fine-tuning strategies will also vary. In \cite{fan21_interspeech}, Fan et al. explored leveraging the wav2vec 2.0 \cite{baevski2020wav2vec} pre-trained model in the speaker verification and language identification task for the first time, where the authors add a pooling layer and a linear transformation on the top of the wav2vec 2.0 model to get the fix-dimensional embedding containing speaker and language information. Vaessen et al. 
\cite{vaessen2022fine} further evaluate the wav2vec 2.0 model on the speaker verification task by comparing different pooling methods and loss functions. However, the results in \cite{vaessen2022fine} still lag behind the well-known ECAPA-TDNN \cite{desplanques2020ecapa} network for speaker verification task. To further enhance the pre-trained model's performance on the speaker verification task, Chen et al. \cite{chen2022large} directly replaced the input of the ECAPA-TDNN network with the weighted summed representation from all the layers of the pre-trained model and achieved excellent performance. The above-mentioned methods always finetuned the whole pre-trained model, which is costly to store a separate task-related copy of the fine-tuned
model parameters for each downstream task. To mitigate this issue, Peng et al. \cite{peng2023parameter} proposed a parameter-efficient fine-tuning strategy by only updating some lightweight adapters and achieved pretty good results. 
Moreover, Cai et al. \cite{cai2023pretraining} demonstrates that, distinct from employing generic pre-trained models, utilizing a Conformer pre-trained specifically on Automatic Speech Recognition (ASR) tasks can enhance performance in speaker verification tasks. They discovered that this ASR-specific pre-training helps alleviate the Conformer model's tendency to overfit during speaker recognition training.

\subsection{From Uni-modality to Multi-modality and Cross-modality}
\label{sec:modality}
One important application of speaker representation learning is to develop an effective and accurate system for personal identity verification. In addition to speech, the human face serves as another critical biometric characteristic for verification tasks. This raises intriguing questions: Is there a connection between information from these two modalities? Do they complement each other? Such questions are increasingly attracting researchers' attention.

\subsubsection{Multi-Modal Person Verification}
Decades ago, researchers identified the complementary nature of information between audio and visual modalities in person verification tasks. They employed a simple score fusion strategy at the decision level~\cite{audiovisualjoint1,audiovisualjoint2,audiovisualjoint3,audiovisualjoint4,audiovisualjoint5} to integrate information from multiple modalities. With the advent of deep learning, the integration of multi-modal information has evolved. Shon et al.~\cite{shon2019noise} introduced various strategies for fusing embeddings from speech and facial images. Following this, Chen et al.~\cite{chen2020multi,qian2021audio} enhanced these approaches with robust backbones and introduced a Noise Distribution Matching (NDM) strategy for better generalization in modality-missing scenarios. Hormann et al.~\cite{hormann2020attention} explored information fusion from different modalities using a multi-scale strategy, rather than merely at the embedding level. Moreover, Sun et al.~\cite{sun2023method} and Liu et al.~\cite{liu2023cross} suggested enhancing the information from one modality with data from another before fusion, aiming for superior results. Additionally, Abdrakhmanova et al.~\cite{abdrakhmanova2023multimodal} introduced thermal modality to person verification, demonstrating enhanced system robustness against data corruption in some modalities.

\subsubsection{Cross-Modal Matching}
The audio and visual modalities not only offer complementary information but also exhibit correlation. Efforts have been made to uncover this correlation by mapping the hidden representations from both modalities to a shared latent space. This existence of correlation has been confirmed by multiple researchers~\cite{nawaz2019deep,horiguchi2018face,kim2019learning,nagrani2018seeing,nagrani2018learnable,wen2018disjoint}, with Nagrani et al.~\cite{nagrani2018seeing} showing that neural networks might even surpass human capabilities in cross-modal face and audio matching tasks. Interestingly, learning the association between face and audio modalities has been found to also enhance single-modality speaker verification performance, as discovered by Shon et al.~\cite{shon2020multimodal} and Tao et al.~\cite{tao20b_interspeech}

\subsubsection{Cross-Modal Knowledge Distillation}
It has been consistently shown that multi-modal systems outperform their single-modality counterparts, and systems based on visual modality particularly excel in identity verification than speech-based systems. However, there are scenarios where only speech is available for verification, such as in voice assistants. To address this, Zhang et al.~\cite{zhang2021knowledge} proposed employing knowledge distillation from multi-modal to single-modal systems across three different levels to enhance performance. Moreover, Jin et al.~\cite{jin2023cross} considered the face recognition model as a teacher, transferring its discriminative capabilities to the speaker recognition model. Furthermore, Tao et al.~\cite{tao2023speaker} utilized a strong face recognition system to identify challenging and noisy-label samples in the audio training set, thereby improving audio-based speaker recognition.

\subsubsection{Boosting Self-supervised Training with Multi-Modal Information}
The complementary capabilities of audio and visual modalities are beneficial not only in supervised systems but also in self-supervised learning. Shi et al.~\cite{shi22c_interspeech} initially pre-trained an audio-visual system with unlabeled speech and lip-region image inputs, then fine-tuned it for downstream audio-visual or audio-only speaker verification tasks. This approach significantly improved label efficiency and the noise robustness of the system. Tao et al.~\cite{tao2023self} used the visual modality to explore a wider range of positive pairs, enhancing the results of contrastive-learning based self-supervised training. Additionally, Cai et al.~\cite{cai2022incorporating} and Han et al.~\cite{han2023self} employed the visual modality to generate more accurate pseudo labels in self-supervised speaker representation learning, underlining the importance of pseudo label quality in achieving optimal system performance.


\section{Robustness and Efficiency}
\label{sec:robust_eff}

\subsection{Robustness: Dealing with Variations}
\label{sec:robustness}


Beyond the conventional definition of robustness, which primarily addresses environmental robustness, we propose to examine robustness from various mismatch perspectives, that include text mismatch~\cite{wang2019usage, yang2020text, tawara2020frame}, language mismatch~\cite{xia2019cross, bhattacharya2019adapting, rohdin2019speaker, thienpondt2020cross}, acoustic environment mismatch~\cite{ma2007effects,qin2022ff} and device/channel mismatch~\cite{chen2020channel}. Notably, the issue of text mismatch~\cite{yang2020text} effectively incorporates both traditional text-dependent and text-independent speaker verification and classification problems.

\begin{figure}[ht!]
  \centering
  \includegraphics[width=0.45\textwidth]{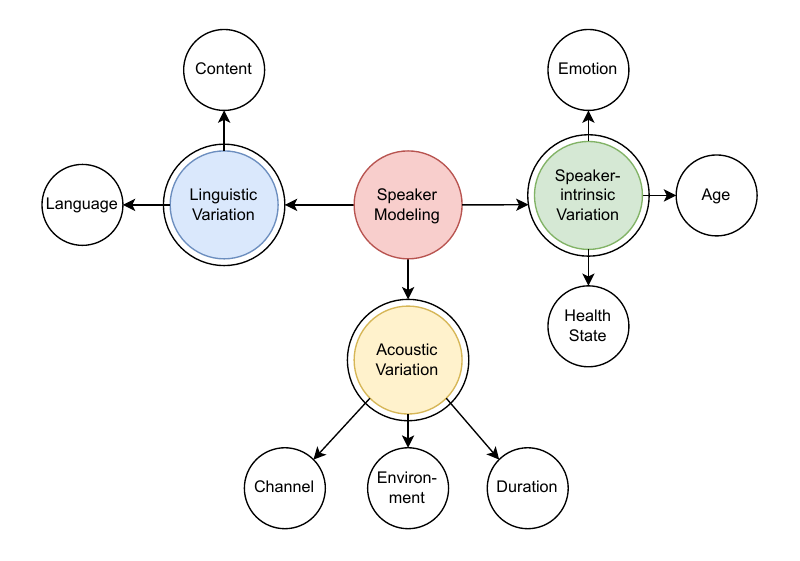}
  \caption{Challenges for robust speaker representation learning}
  \label{fig:rob_eff}
\vspace{-10pt}
\end{figure}

In tackling the robustness challenges arising from various factors in speaker representation learning, researchers often employ similar research ideas and strategies. The primary strategies can be summarized in two main aspects:
\begin{enumerate}[leftmargin=4mm]
    \item Firstly, researchers aim to enhance the model's robustness against intra-class variation. This is typically accomplished by enhancing the model's capacity to discern the timbre characteristics of a speaker across varying conditions, ensuring that the model can produce a more cohesive and consistent speaker modeling space, regardless of environmental fluctuations or changes in the speaker's intrinsic states.
    \item Secondly, techniques akin to adversarial training are engaged to lessen the influence of extraneous information. Within these approaches, the model is trained to detect and neutralize variables unrelated to speaker identity features, such as ambient noise, channel artifacts, emotional states, health conditions, and other factors, thereby rendering the speaker identity features more salient and distinct.
    
\end{enumerate}
This section provides a detailed review of various types of nuisance information and explores the specific challenges and solutions they pose to speaker representation learning.

\subsubsection{Text Dependency}
\label{sec:text}
Traditional speaker recognition tasks are typically divided into two categories: text-dependent and text-independent. In text-dependent scenarios, the speaker's identity and the content of their test speech must align with those of the enrollment speech. Conversely, text-independent verification does not impose any constraints on the spoken content. The majority of application scenarios and integrated speech-related tasks, such as speaker extraction, necessitate that the speaker representation derived from the enrollment speech be minimally influenced by the text content, adhering to the principle of ``consistency" as previously mentioned.

Although standard training loss functions can mitigate the variability introduced by text information to a certain degree—for instance, softmax-based loss functions guide the learned representations to approximate the weight vectors associated with their respective classes in angular space, while center loss further enforces the distinctiveness of representations for different utterances by the same speaker—numerous studies have shown that speaker embeddings trained with these methods still retain the capability to distinguish content~\cite{wang2017does}.

Hong et al.~\cite{hong2023decomposition} tackled this issue by minimizing frame-level feature discrepancies, thereby indirectly eliminating phoneme-related information. In contrast, Wang et al.~\cite{wang2019usage}, Zhou et al.~\cite{zhou2019cnn}, and Chen et al.~\cite{chen2021phoneme} initially incorporated phoneme information into the speaker encoder, enabling the model to identify such details, before removing it to minimize the impact of phoneme information on feature representation. Recognizing the relative simplicity of text-dependent tasks, Wang et al.~\cite{wang2022novel} introduced a novel approach that re-segments speech based on phoneme categories and consolidates frame-level representations within these categories. This technique aligns the verification process with speaker embeddings from the same phoneme category, effectively transforming a text-independent task into a text-dependent one. Furthermore, Yang et al.~\cite{yang2020text} explored a method to separate speaker information from content information, granting the system the flexibility to decide whether to account for the text information of the current input or to disregard it. The proposed Speaker-text factorization network (STFNet) presented enhanced performance in text-independent, text-dependent, and text-adaptive tasks across various text mismatch conditions. Similarly, the work by Liu et al.~\cite{liu2024disentangling} demonstrated that separating speaker information from content information in text-independent verification tasks yields more robust speaker embeddings.

In summary, the process of decoupling content information from speaker representations primarily utilizes a strategy of branch restriction and combination reconstruction. This involves designing distinct branches for speaker and content information, where each branch models its respective information and is governed by constraint functions related to that information. To ensure that this process does not lose excessive information, the system often merges the outputs of both branches later on, aiming to reconstruct and fit the original undecoupled information.

\subsubsection{Cross-Language Robustness}
\label{sec:language}

Although speaker embedding extractors are designed to extract speaker identity information from audio, the spoken language can still impact the system's performance. Addressing this challenge, adversarial training emerges as an effective strategy to bolster language robustness of the system~\cite{li2021gradient,chen2020adversarial}. Typically, the adversarial is conducted at the utterance level. Lin et al.\cite{lin2020framework} discovered that reducing language discrepancies at both the frame and utterance levels significantly outperforms the traditional utterance-level approach. They also found that employing distinct batch normalization for data in different languages yields better results. The goal of adversarial training is to align speech data from various languages into a cohesive distribution. Nevertheless, when data from the target language lacks labels, relying solely on distribution information proves insufficient. To address this, Chen et al.\cite{chen2021self}, Li et al.\cite{li22m_interspeech}, and Mao et al.\cite{mao2023cluster} suggest a hybrid approach that combines supervised training with source language data and self-supervised training with target language data to enhance performance. Furthermore, Hu et al.~\cite{hu2022class} advance self-supervised training-based methods by focusing on the alignment of within-class and between-class distributions.

\subsubsection{Acoustic Channel Variability}
\label{sec:acoustic}
The effectiveness of speaker verification systems is significantly influenced by complex application acoustic environments. Therefore, mitigating the impact of external acoustic conditions is paramount for these systems. A prevalent strategy involves employing adversarial training~\cite{chen2020channel,zhou2019training,meng2019adversarial} to eliminate acoustic environment or channel information, necessitating additional channel labels during the adversarial training phase. Advancing this approach, Peri et al.\cite{peri2020robust} utilized unsupervised adversarial invariance training to separate speaker-discriminative information from all other information present in the audio recordings without supervising acoustic conditions. Similarly, Luu et al.\cite{luu2020channel} applied adversarial training to minimize channel discrepancies between audio segments from the same speaker but different recordings.
Furthermore, Li et al.\cite{li2021gradient} introduced a method to synchronize the gradients between noisy utterances and their clean versions, aiming to reduce the adverse effects of external noise. In addition to filtering out external acoustic environment information, Gu et al.\cite{gu2023dynamic} developed a dynamic convolutional neural network capable of automatically adjusting its parameters in response to various acoustic environments.

\subsubsection{Intra-Speaker Variability}
\label{sec:speaker}
In addition to the domain mismatch problems caused by environmental factors, changes in a speaker's personal state also pose challenges to speaker timbre modeling. Timbre is a characteristic of the speech signal that varies with internal factors such as the speaker's age, emotion, and health status. During the process of learning speaker representations, these factors significantly increase the intra-class variance. Therefore, some researchers have specifically studied these factors in the hope of constructing speaker representations that are robust to these variations.

\noindent \textbf{Age:}
Cross-age speaker modeling holds significant practical importance, such as identifying suspects from old telephony scam recordings or unlocking voice recognition systems registered many years ago. Despite its importance, research in this area is limited due to the scarcity of relevant data.

Previous research works have pointed out that speaker identification across age presents clear challenges: in \cite{kelly2011effects, kelly2012speaker}, researchers found that age has a more significant impact on target (same speaker) scores than non-target (different speakers) scores, significantly lowering the scores for target tests and thereby increasing the system's equal error rate (EER). In \cite{kelly2016score}, researchers proposed a score calibration method that alleviates the negative impact on the speaker recognition systems in cross-age scenarios to some extent. The latest research published in \cite{qin2022cross} presents an explicit model architecture to weaken the encoding of age information in speaker representations. Specifically, an Age Decoupling Adversarial Learning (ADAL) module is proposed, which uses an attention mechanism to extract age-related information from high-dimensional feature maps, achieving decoupling of the age component from the speaker identity component. The purpose of this method is to minimize the encoding of age information in speaker representations, thereby improving the model's robustness to age variability.

\noindent \textbf{Emotion:}
When speakers are in different emotional states, their pronunciation characteristics change, and emotional states can significantly affect the pitch, rhythm, and intensity of speech. For example, when happy or surprised, one's voice tends to be sharper, while it becomes lower when sad. These emotion-induced changes pose a significant challenge to speaker modeling, resulting in a substantial decline in speaker recognition performance across emotional states \cite{parthasarathy2017study, pappagari2020x}.

In the paper \cite{li2020segment}, Kai et al. addressed this issue by proposing an emotion-adversarial learning approach that explicitly suppresses emotional information by introducing a Gradient Reversal Layer (GRL), aiming to learn more robust and emotion-independent speaker features. Another study \cite{lertpetchpun2023instance} adopted a Temporal Normalization Layer (TNL) to reduce the model's sensitivity to emotion, thereby increasing recognition accuracy in emotion-mismatched environments.
In recent research \cite{tian2024learning}, to further enhance the performance of cross-emotion speech recognition, researchers employed the CopyPaste technique for data augmentation, creating more parallel emotional corpora. Additionally, they proposed a new training criterion that explicitly minimizes the correlation between speaker representations and emotional information.

\noindent \textbf{Health:} Additionally, the speakers' physical condition can affect the stability of voiceprint extraction. For example, a person's voice when he has a cold is noticeably different from his voice in normal health; this is also termed ``cold-affected speech'' \cite{tull1996analysis}. Chen et al. \cite{chen2021health} demonstrated a robust speaker verification system that supports mask-wearing during the COVID-19 pandemic. Another example is speech issues caused by pathological reasons, such as a decline in vocabulary richness and speech fluency resulting from cognitive impairments like Alzheimer's disease (AD) \cite{ehghaghi2023factors}.

\subsubsection{Duration Robustness}
\label{sec:duration}
The accuracy in speaker representation extraction from short-term speech has always been a hot research topic. Compared to speech lasting several seconds or even tens of seconds, short speech (less than 1 second) contains very limited information and is easily affected by other variations, such as being dominated by content or channel information. In real-world scenarios, users are often required to record longer registration speech in a cooperative situation, but the length of test speech during application tends to be more flexible, resulting in some very short sentences.

To enhance the robustness of speaker representation learning on short-term speech, researchers typically explore from the following perspectives:

\begin{enumerate}[leftmargin=3mm,label=\roman*.]
    \item \textit{Design of more efficient aggregation methods}: Typically, the pooling (aggregation) function utilizes single-scale features from the final frame-level layers. However, research indicates that incorporating multi-scale features can be advantageous. For instance, multi-scale feature aggregation (MSA) can aggregate speaker information across different timescales and layers, which has been proven effective for short-duration utterances~\cite{hajavi2019deep,tang2019deep,gao2019improving,jung2020improving}.
    \item \textit{Explicit alignment of short and long speech representations}: By designing special models and loss functions to make the embeddings of short-term and long-term speech as close as possible. For example, metric-based loss can be used to train models so that the features of short and long speech from the same speaker are more closely packed in the embedding space~\cite{kye2020meta}. Similarly, adversarial training~\cite{liu2020text} and teacher-student transfer learning~\cite{jung2019short,sang2020open}  can be leveraged to achieve this objective.
    \item \textit{Prediction from short-term embeddings to long-term embeddings}: Different from the direct alignment of long and short speech representations, this method does not directly restrict the two to be similar, but aims to process a given short speech embedding through a neural network model, mapping it to the long speech embedding space. This is achieved through a mapping or transformation network~\cite{guo2017cnn,zhang2018vector} that learns the relationship between long and short speech.
\end{enumerate}
\subsubsection{Learning from Noisy Labels}
\label{sec:noisy_label}
Labeling large-scale datasets is a costly process. Many speaker verification datasets are collected automatically~\cite{nagrani2020voxceleb,li2022cn}, potentially introducing labeling errors. It is crucial to detect noisy labels and either discard or correct them to enhance the system's performance. Qin et al.\cite{qin2022simple} utilized an iterative strategy for noisy label detection to consistently identify and address noisy labels, while Tong et al.\cite{tong2021automatic} introduced a noise correction loss to rectify noisy labels during training. Given that the speaker verification model comprises a front-end embedding extractor and a scoring backend, Borgstrom et al.\cite{borgstrom2020bayesian} and Li et al.\cite{li2022speaker} explored label correction at different stages of the speaker embedding extraction model. Additionally, multi-modal systems generally outperform speech-based single-modality systems. Tao et al.\cite{tao2023speaker} suggested utilizing a robust audio-visual system to identify noisy labels within the audio dataset. Although the proportion of noisy labels in genuinely collected data is relatively small, self-supervised learning often generates pseudo-labels through pre-trained models, which significantly contain noisy labels, adversely affecting the system's performance\cite{fathan2022impact}. Chen et al.\cite{chen2023unsupervised} recommended a combined online and offline label correction strategy for using pseudo-labels. Cai et al.\cite{cai2021iterative} identified noisy labels based on clustering confidence. Moreover, Tao et al.\cite{tao2022self} proposed distinguishing between noisy and clean labels based on loss values during training, a method further refined by Han et al.\cite{han22b_interspeech} through a dynamic loss-gate strategy.

\subsection{Efficiency: Model Compactness and Inference Speedup} 
\label{sec:efficiency}
Efficiency is an important research direction for making technologies deployed in the real-world products, especially for the on-device applications. In this section, we will discuss several key aspects related to efficiency on modeling speaker information, including quantization~\cite{zhu2021binary, wang2023lowbit}, knowledge distillation~\cite{wang2019knowledge, peng2022label, liu2023distilling}, efficient architecture design~\cite{xue2014singular, ko2020prototypical, nunes2020mobilenet1d, liu2023depth}, and policy design tailored for specific tasks~\cite{li2022towards}.

\subsubsection{Model Quantization}
\label{sec:quantization}
Speaker encoders typically utilize float-32 for parameter representation, which introduces unnecessary redundancy. Wang et al.~\cite{wang23u_interspeech} demonstrated that a 4-bit quantized version of ResNet34 achieved performance comparable to its float-32 counterpart through a K-means-based quantization method. Further studies by Wang et al.~\cite{wang2023lowbit} and Li et al.~\cite{li23t_interspeech} have analyzed the impact of quantization across different systems and components. Moreover, Zhu et al.~\cite{zhu21_interspeech} and Liu et al.~\cite{liu23g_interspeech} investigated the application of extremely low bit, i.e. binary quantization, for the speaker verification task, revealing that even with severe quantization, models can maintain robust performance on the VoxCeleb dataset.
\subsubsection{Knowledge Distillation}
\label{sec:kd}
\begin{figure}[ht!]
  \centering
  \includegraphics[width=0.46\textwidth]{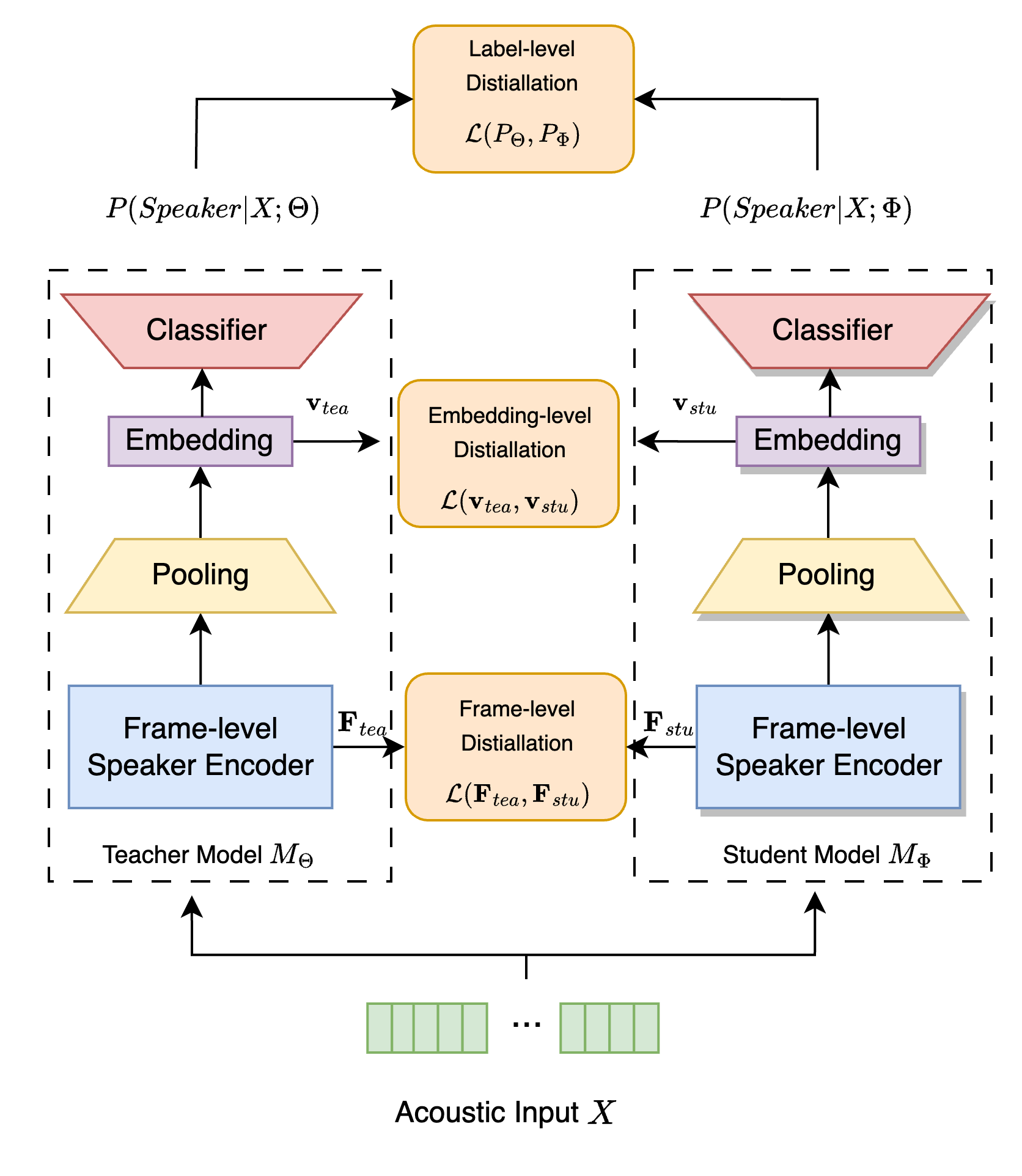}
  \caption{Knowledge distillation at different levels}
  \label{fig:distill}
\end{figure}

Knowledge distillation (KD) offers an alternative approach for creating lightweight models without sacrificing performance. As shown in Fig.~\ref{fig:distill}, KD can be applied at different levels. Wang et al.~\cite{wang2019knowledge} introduced both label-level and embedding-level KD to minimize the performance differential between larger and smaller models. To exploit unlabeled data, Peng et al.~\cite{peng2022label} developed a label-free KD technique within a self-supervised learning framework. These methods primarily focus on utterance-level knowledge transfer. Additionally, Liu et al.~\cite{liu2023distilling} and Liu et al.~\cite{liu2022self} advanced the field by proposing a multi-level distillation strategy, further enhancing model performance.  In pursuit of high-performance, lightweight models, Cai et al.~\cite{cai22_interspeech} combined improved architectural designs with knowledge distillation. Besides, Truong et al.~\cite{truong2024emphasized} found that emphasizing the classification probabilities of non-target speakers during knowledge distillation can improve the system performance further. 
All the aforementioned studies utilized a pre-trained teacher model for knowledge distillation. To obviate the necessity for such a model, Liu et al.~\cite{liu2022self} introduced a self-teacher approach, where the teacher model is jointly trained alongside the student model.
\subsubsection{Efficient Architecture Design}
\label{sec:effarch}
Direct architectural modifications also serve as a fundamental strategy for efficiency improvement. Several studies~\cite{cai22_interspeech,nunes2020mobilenet1d,ko2020prototypical,wu2022rsknet,chen2023lightweight,li2023lightweight,wang2023lightweight} have adopted efficient modules within CNN-based speaker encoders. Wang et al.~\cite{wang2022efficienttdnn} differentiated themselves by employing a neural architecture search (NAS) to identify compact and efficient networks. Similarly, Liu et al.~\cite{liu2022df,liu2023depth} observed superior performance in deeper, narrower networks compared to their wider, equivalently-sized counterparts, leading to the proposition of a depth-first network design strategy.
\subsubsection{Efficient Policy Design}
\label{sec:effpolicy}
The above methods are all task-agnostic, general-purpose approaches. For the speaker verification task specifically, enrollment is conducted only once, but evaluation occurs as frequently as the system is used. Considering this characteristic, Li et al.~\cite{li2022towards} proposed an asymmetric enroll-verify structure for speaker verification. A large model is used during enrollment, while a smaller model is employed for verification. Similar approaches are further investigated in ~\cite{gao2024post}.

\section{Explainabilty}
\label{sec:explain}
As speaker modeling advances, researchers increasingly focus on exploring its underlying mechanisms, particularly with an emphasis on interpretability. In the following section, we will introduce the commonly adopted methods for research on the interpretability of speaker representation learning.




\subsection{Probing Tasks Based Methods}
\label{sec:probe}

\emph{Exploring embedding capacity through probing tasks.} We previously proposed the first analysis framework to quantitively compare the characteristics and capabilities of speaker embeddings by designing various probing tasks \cite{wang2017does}. The effectiveness of different embeddings in modeling various information, such as speaker, text, and environmental information are compared and analyzed. Later, this approach is adopted to probe the capabilites of representations extracted from more advanced models such as TDNN~\cite{raj2019probing} and transformers~\cite{singla2022audio}. Recently, the design philosophy of the widely-adopted SUPERB benchmark~\cite{yang2021superb} is similar in that it aims to analyze the performance of various pretrained models across different downstream tasks (such as speaker recognition, emotion recognition, gender recognition). This is achieved by introducing simple probing heads tailored to different attributes, thereby facilitating the measurement of performance in a structured manner.

 \subsection{Component Ablation}
 \label{sec:ablation}
 \emph{Measuring component importance via ablation}. An ablation study aims to understand the importance of specific components of a model by systematically removing (ablating) parts to determine their contribution to the final performance. This method has been adopted in many studies to build better speaker representation learners, with ablated components including network layers~\cite{zhang2022mfa,aldeneh2024can}, loss functions~\cite{kataria2020analysis,kim2019deep}, and training strategies~\cite{thienpondt2021idlab,zhang2023adaptive}. For instance, to analyze the capabilities of different self-supervised learning (SSL) models and the impact of the downstream model, the study~\cite{aldeneh2024can} employs an ablation strategy where components are removed step-by-step. The works~\cite{thienpondt2021idlab,zhang2023adaptive} validate the strategy of large-margin fine-tuning by comparing systems with and without its application. Overall, component ablation is a widely utilized methodology in numerous research studies.

\subsection{Relevance Analysis}
\label{sec:relevance}
 \emph{Understanding intermediate layers through relevance analysis methods.}
Interpreting intermediate representations instead of the final embeddings is also very meaningful, as it can help us comprehend the changing trends of different attributes during the cross-layer flow of information.
~\cite{ashihara2024self} adopted layer-wise similarity analysis based on linear centered kernel alignment (LinCKA) to measure layer-wise similarity of different speaker encoders. Similar relevance analysis methods such as canonical correlation analysis (CCA) were adopted for the analysis of different pretrained speech models~\cite{pasad2023comparative}. 

\subsection{Visualization Methods}
\label{sec:visualize}
 \emph{Measuring the importance through visualization.} Researchers proposed using a visual analysis solution, Class Activation Map (CAM), in image visualization to examine data augmentation schemes during speaker representation learning \cite{Jiang2021LayerCAMEH, Li2023VisualizingDA}. This approach intuitively reflects the model's robustness to scenarios like noise after data augmentation. Zhang et al.~\cite{zhang2023study,zhang2024speaker} employed attribution algorithms to visualize the importance of features. Similar approaches have been applied previously to compare the effectiveness of Res2Net and ResNet architectures in speaker representation learning \cite{zhou2021resnext}. Moreover, in the task of Speaker Antispoofing \cite{Himawan2019VoicePA}, Himawan et al. utilized CAM analysis and discovered the critical role of high-frequency information in distinguishing genuine speech from forged speech.


\subsection{Explainable Networks}
\label{sec:xai}
\emph{Explainble achitecture design for decision transparency.} The Concept Bottleneck Model (CBM)~\cite{koh2020concept} approach incorporates an intermediate layer specifically designed to capture representations related to human-understandable concepts or attributes. This structure allows the model's decision-making process to be linked directly to specific concepts, offering a transparent explanation pathway for each prediction. This approach has been successfully applied to the task of speaker verification, as detailed in~\cite{wu2024explainable}, where it considers attributes such as gender, nationality, age, and profession.

\section{Applications and Integration Strategies}

As mentioned earlier, speaker representation plays an important role in many related tasks, providing critical timbre information for the main task. This involves selecting appropriate speaker representations and effectively integrating this information into the main task. In this section, we will first provide a general overview of the methods used to integrate speaker information. Following this, we will delve into classic examples of how these methods are applied across different downstream tasks\footnote{Note that our aim is not to cover all related papers, as this is not feasible. Instead, we will focus more on introducing common practices and example publications in the field. }.  
\label{sec:application}

\subsection{Integration Strategies}
\label{sec:integration}

Generally speaking, strategies for integrating speaker information into other tasks can be mainly divided into two categories: off-the-the usage and joint optimization. Regardless of the strategy, we need specific fusion methods to integrate speaker information into other tasks. Additionally, both off-the-shelf usage and joint optimization have their advantages and disadvantages, and the choice between them should be based on specific needs during implementation.

\subsubsection{Off-the-Shelf Usage}
\label{sec:offtheshelf}

Firstly, a common practice is to use speaker embedding models that have been pre-trained on speaker recognition tasks. This method treats the pre-trained model as an additional preprocessing component for extracting speaker embeddings~\cite{medennikov2020target,medennikov2020stc,ttscooper2020zero} and integrates them directly into the target system. These speaker embeddings can serve as input features~\cite{cheng2022target} for the model, enriching it with information about the speaker's identity; similarly, they can also be part of the target output. By designing specific loss functions, the model's output can be constrained and guided to ensure the speech retains the correct speaker characteristics.
\subsubsection{Joint Optimization}
\label{sec:joint}
Another method of integrating speaker information is to jointly train the speaker model with the target task module. This strategy does not simply use pre-trained speaker representations but learns speaker characteristics simultaneously with the target task. This end-to-end training approach allows for mutual enhancement and joint optimization~\cite{ji2020speaker,kim2021conditional} between speaker recognition and the target task. In this approach, the speaker modeling network and the main task network (such as speech recognition, speech synthesis, etc.) share some of the network structure, while their respective task-specific layers are responsible for capturing the detailed features of their domains. Then, an appropriate fusion method is used to integrate the information from different branches.

\subsubsection{Speaker Information Fusion Methods} When speaker information is used as input or extra conditions in both pretrained and jointly optimized frameworks, a key challenge is how to effectively integrate this information into the main network of the target task. There are several common methods to achieve this, including non-parameterized methods such as concatenation and feature addition, as well as parameterized conditioning schemes such as Feature-wise Linear Modulation (FiLM)~\cite{perez2018film}, Conditional Layer Normalization (CLN)~\cite{chen2021adaspeech}, and Adaptive Instance Normalization (AdaIN)~\cite{huang2017arbitrary}, etc. From a unified perspective, AdaIN, CLN, and FiLM can all be viewed as methods that use conditional information to generate scaling and shifting parameters, thereby modulating feature normalization or linear transformation. This approach enables neural networks to dynamically adapt to different conditional speaker information. 



\subsubsection{Strategy Comparison}
Both strategies are widely used, each with its own advantages and limitations:

The pre-trained speaker representations can simplify the target task design, avoiding additional architectural design and optimization strategies. The modular design is flexible and can leverage existing models, often trained on large-scale data, yielding high-quality, generalized speaker representations. However, pre-trained models may perform worse on specific tasks and datasets due to lack of task-specific optimization and adaptation to new data distributions. Efficient fusion algorithms are needed to fully utilize the highly compressed representations. Performance evaluation typically relies on speaker verification tasks, but this does not always correlate with performance on various downstream tasks~\cite{jung2023search}, complicating model selection.

The joint trained framework adapts better to specific tasks and datasets, reducing dependence on the fusion module. However, it increases model optimization complexity, requiring careful tuning. Limited training data for the main task can lead to overfitting, reducing generality and performance on unseen speakers or data. The rise of large models and massive datasets has mitigated this limitation to some extent.

\subsection{Speaker Diarization}

\label{sec:diarization}
Speaker diarization can be classified into cluster-based and end-to-end neural approaches. The former primarily relies on pretrained speaker embeddings, while the latter often opts for joint training.

\subsubsection{Clustering Based Diarization System}
A clustering-based diarization system~\cite{shum2013unsupervised,sell2014speaker} typically involves three stages: (1) chunking the audio into short segments, (2) extracting speaker embeddings for each segment, and (3) clustering the segments based on the extracted speaker embeddings. In the second stage, any pre-trained speaker encoder can be utilized to extract the speaker embeddings.

\subsubsection{Neural Diarization System}
The neural diarization (ND) system, such as an end-to-end system~\cite{fujita2019end_lstm,fujita2019end_sa,horiguchi2022encoder,horiguchi2020end,takashima2021end,chen23n_interspeech,chen2024attention} or a TS-VAD~\cite{medennikov2020target,medennikov2020stc,cheng2022multi,wang2022target,wang2022incorporating,cheng2022target} system, typically takes acoustic features like MFCCs as input. The system implicitly learns speaker representations guided by diarization objectives. Additionally, the acoustic features can be replaced by frame-level speaker representations~\cite{cheng2022target} extracted from a pre-trained speaker encoder, which contains more speaker-related information. In the TS-VAD system, the speaker representations can also serve as enrollment information, indicating the target speaker's identity corresponding to the system's output. Besides, recently proposed prompt-based diarization uses learnable vectors \cite{jiang2024prompt} to represent certain speaker-specific characteristics or utilizes multimodal prompts~\cite{jiang2024target} to associate with speaker attributes.

\subsection{Speech Recognition}
\label{sec:asr}
In traditional speech recognition, speaker information is commonly utilized in two ways:
    \begin{itemize}[leftmargin=3mm]
        \item Target Speaker Adaptation: Speech recognition systems can personalize adaptation based on different speakers' speech characteristics to enhance recognition accuracy. Relevant work often involves concatenating a speaker representation extracted from a pre-trained model into the input of the speech recognition model. Various models like \textit{i}-vector~\cite{gupta2014vector}, x-vector~\cite{geng2022speaker}, among others, have been explored by researchers aiming at tailor systems for target individuals to achieve improved performance.
        \item Constructing a speaker-agnostic, universally applicable speech model by explicitly removing speaker information.  Meng et al. proposed using speaker adversarial training to obtain more robust speech recognition models across different speakers~\cite{meng2018speaker}.
    \end{itemize}

Recent works~\cite{delcroix2019end,moriya2023streaming} introduce speaker information to directly recognize the speech content of a specific person in the cocktail party problem, known as target speech recognition. This can be considered another application of speaker information in speech recognition.

\subsection{Speech Generation}
For the speech generation task, where the voice of a target speaker needs to be generated based on input text or source speech, a timbre identifier should be provided to guide the direction of the generation.

\label{sec:generation}
\subsubsection{Speech Synthesis} Speaker information modeling can be used to generate synthesized speech with specific speaker characteristics. The common approach directly integrates pretrained representations alongside text into a multispeaker speech synthesis system \cite{ttsjia2018transfer, ttschen2019cross, ttscooper2020zero}, ensuring that the synthesized speech retains the vocal features of a particular speaker.
In addition to pretrained global speaker embeddings, \cite{ttszhou2022content} employs jointly trained local speaker embeddings to achieve finer control over speaker identity. Moreover, beyond serving as a condition for model prediction, speaker identity information can also act as a target for speech generation, providing constrained guidance. For instance, \cite{ttscai2020speaker} introduces a cycle loss related to speaker embeddings to explicitly constrain the speaker identity of synthesized speech.

In more recent synthesis architectures such as the prompt-based systems~\cite{wang2023neural,du2024unicats,du2024vall}, there is no explicit ``speaker embedding" involved. Instead, the acoustic prompt essentially models the speaker information and acoustic information together. This approach shows more promising results in the preservation of speaker identity.

\subsubsection{Voice Conversion} The goal of voice conversion is to transform one speaker's voice into another's while preserving the speech content~\cite{sisman2020overview}. Similar to TTS, modeling speaker information is crucial for achieving high-quality voice conversion. One-hot representation is commonly used to represent the target speaker's voice, yet this method lacks flexibility in expanding the target speaker set. Liu et al. proposed integrating speaker statistics modeling into voice conversion systems~\cite{liu2021non}, achieving notable performance improvements in voice similarity. 
    
\subsection{Target Speaker Frond-End processing}
\label{sec:tse}

Speech front-end processing algorithms have a wide range of applications, including tasks such as speech enhancement, voice activity detection, and speech separation. Sometimes, we want to focus specifically on front-end processing algorithms for a particular speaker, which has led to the development of a series of speaker-dependent front-end processing algorithms such as personal voice activity detection~\cite{ding2019personal,ding2022personal}, target speaker enhancement, and target speaker extraction~\cite{zmolikova2023neural,wang2024wesep}. These methods often specify the target speaker by introducing additional speaker encoding, which can be either pretrained~\cite{vzmolikova2019speakerbeam,xu2020spex,ge2020spex+,hao4611108xtfgridnet} or jointly trained~\cite{wang2018voicefilter,yu2023tspeech,liu2023xsepformer}.

\section{Benchmark Datasets and Open-Source Toolkits}
\label{sec:open}

In recent years, we have seen the release of several new datasets such as VoxCeleb~\cite{nagrani2020voxceleb}, CNCeleb~\cite{li2022cn}, and 3D-Speaker~\cite{zheng20233d}. Alongside these, numerous open-source tools~\cite{watanabe2018espnet, yao2021wenet, ravanelli2021speechbrain, tong2021asv, wang2023wespeaker} dedicated to speaker representation learning have emerged, including the WeSpeaker project initiated by ourselves~\cite{wang2023wespeaker,wang2024advancing}. In this section, we will provide a summarization of these popular datasets and open-source toolkits. 

\begin{table*}[ht!]
\footnotesize
\centering
\caption{{Datasets for Speaker Representation Learning.} TI and TD denote text-independent and text-dependent, respectively.
}
\begin{adjustbox}{width=.99\textwidth,center}
\begin{threeparttable}
\begin{tabular}{llllllrl}
\toprule

Name                           & Year      & Language         & Data Source                                & TI/TD & Spk \# & Utt \#    & Duration (hrs) \\
\hline
TIMIT~\cite{fisher1986darpa,zue1996transcription}                          & 1986      & English          & telephone                                  & TI    & 630    & 6,300     & -              \\
SWB~\cite{godfrey1992switchboard}                            & 1992      & English          & telephone                                  & TI    & 3,114  & 33,039    & -              \\
NIST SRE~\cite{gonzalez2014evaluating,greenberg2020two}                       & 1996-2020 & Multilingual     & telephone, microphone                      & TI    & -      & -         & -              \\
RSR2015~\cite{larcher2014text}                        & 2015      & English          & mobile, tablet                             & TD    & 300    & 190,000   & -              \\
RedDots~\cite{lee2015reddots}                        & 2015      & Multilingual     & mobile                                     & TD    & 62     & 13,500    & -              \\
Librispeech~\cite{panayotov2015librispeech}                    & 2015      & English          & Reading Audiobooks                         & TI    & 2,484  & 292,367   & 982            \\
SITW~\cite{mclaren2016speakers}                           & 2016      & English          & open-source media                          & TI    & 299    & 2,800     & -              \\
Aishell-1~\cite{bu2017aishell}                      & 2017      & Chinese          & mobile                                     & TI    & 400    & 140,000   & 500            \\
VoxCeleb1~\cite{nagrani17_interspeech}                      & 2017      & Mostly English   & YouTube                                    & TI    & 1,251  & 153,516   & 351            \\
VoxCeleb2~\cite{chung18b_interspeech}                      & 2018      & Multilingual     & YouTube                                    & TI    & 6,112  & 1,128,246 & 2,442          \\
Aishell-2~\cite{du2018aishell}                      & 2018      & Chinese          & mobile                                     & TI    & 1,991  & 1,000,000 & 1,000          \\
VOICES~\cite{richey18_interspeech,nandwana2019voices}                         & 2018      & English          & Far-field microphones (noisy room)         & TI    & 300    & 374,688   & 1,440          \\
librilight~\cite{kahn2020libri}                     & 2019      & English          & Reading Audiobooks                         & TI    & 7,439  & 219,041   & 57,706         \\
HI-MIA~\cite{qin2020hi}                         & 2020      & Chinese, English & microphone, mobile                         & TD    & 340    & 3,940,000 & -              \\
BookTubeSpeech~\cite{pham2020toward}                 & 2020      & English          & BookTube (Youtube)                         & TI    & 8,450  & 38,707    & -              \\
CN-Celeb1~\cite{fan2020cn}                      & 2020      & Chinese          & Bilibili                                   & TI    & 1,000  & 130,109   & 274            \\
CN-Celeb2~\cite{li2022cn}                      & 2020      & Chinese          & multi-media sources                        & TI    & 2,000  & 529,485   & 1,090          \\
Multilingual LibriSpeech (MLS)~\cite{pratap20_interspeech} & 2020      & Multilingual     & Reading Audiobooks                         & TI    & 6,332  & -         & 50,834         \\
FFSVC 2020~\cite{qin2020ffsvc}                     & 2020      & Mandarin         & Close-talk cellphone, far-field microphone & TI,TD & -      & -         & -              \\
MULTISV~\cite{movsner2022multisv}                        & 2021      & Mostly English   & Clean speech from Voxceleb                 & TI    & 1,090  & 44,574    & 77             \\
VoxMovies~\cite{brown2021playing}                      & 2021      & Mostly English   & YouTube                                    & TI    & 856    & 8,905     & -              \\
NIST SRE CTS Superset~\cite{sadjadi2021nist}          & 2021      & Multilingual     & telephone                                  & TI    & 7,011  & 605,760   & -              \\
Gigaspeech$^\dagger$~\cite{chen21o_interspeech}                     & 2021      & Multilingual     & Audiobooks/Youtube/Podcast                 & -     & -      & -         & 40,000         \\
Wenetspeech$^\dagger$~\cite{zhang2022wenetspeech}                    & 2022      & Chinese          & Youtube/Podcast                            & -     & -      & -         & 22,435         \\
VoxTube~\cite{yakovlev2023voxtube}                        & 2023      & Multilingual     & YouTube                                    & TI    & 5,040  & 4,439,888 & 4,933          \\
3D-Speaker~\cite{zheng20233d}                     & 2023      & Chinese          & multiple-device                            & TI    & 10,000 & 579,013   & 1,124          \\
VoxBlink~\cite{lin2024voxblink}                       & 2023      & Multilingual     & YouTube                                    & TI    & 38,065 & 1,455,190 & 2,135          \\
VoxBlink-Clean~\cite{lin2024voxblink}                 & 2023      & Multilingual     & YouTube                                    & TI    & 18,381 & 1,028,095 & 1,670          \\
CRYCELEB~\cite{budaghyan2024cryceleb}                       & 2023      & Baby Crying      & mobile (Samsung A10 smartphone)            & TI    & 786    & 26,093    & 7             \\

\bottomrule

\end{tabular}
\begin{tablenotes}\footnotesize
\item $^\dagger$: Datasets without providing speaker labels, which can be used in the self-supervised speaker representation learning.
\end{tablenotes}
\end{threeparttable}
\label{table:dataset_list}
\end{adjustbox}
\end{table*}
\subsection{Dataset}
\label{sec:data}
\begin{figure}[ht!]
  \centering
  \includegraphics[width=0.5\textwidth]{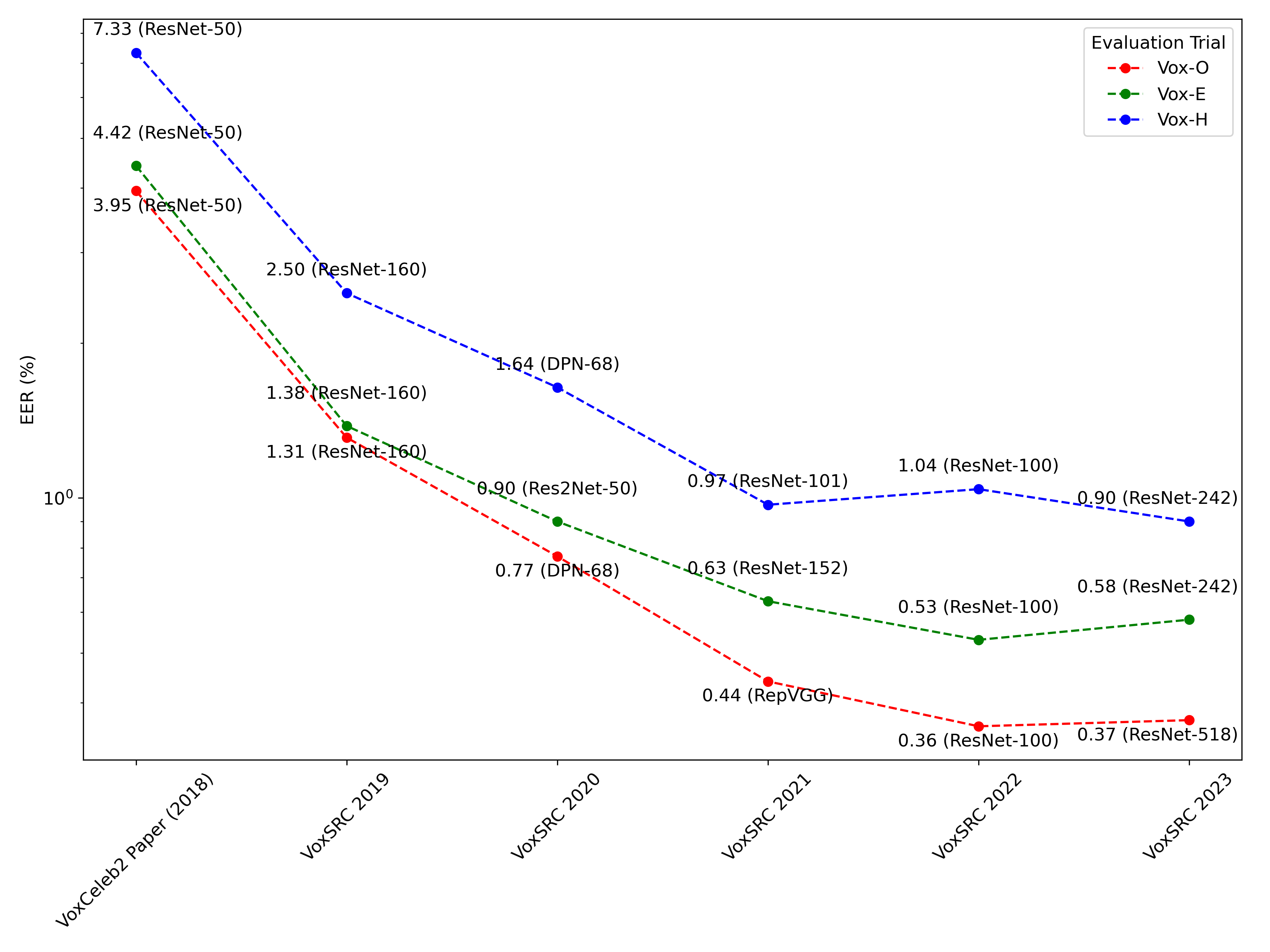}
  \caption{\textbf{Performance trend on Voxceleb dataset.} We have curated results from the original Voxceleb2~\cite{chung18b_interspeech} paper, as well as from the reports of each year's winning system in the VoxSRC~\cite{zeinali2019but,thienpondt2020idlab,zhao2021speakin,makarov2022id,zheng2023unisound}. Additionally, for VoxSRC 2020, we have included results from the report of the second-place winner~\cite{xiang2020xx205}, since the winner~\cite{thienpondt2020idlab} did not provide results for Voxceleb. All the referenced systems were trained using the Voxceleb2 development set, and the VoxSRC selected results are from the best single systems in the report rather than final fusion system.}
  \label{fig:vox_res_trend}
\end{figure}

Research on speaker recognition has been ongoing for many years. Since the late 20th century, the National Institute of Standards and Technology (NIST) in the United States has continuously released related data and held the Speaker Recognition Evaluation Challenge (SRE)~\cite{gonzalez2014evaluating} to advance the progress of research. However, the collection settings for NIST SRE data are quite constrained, and the data is not free. To further promote research in this area, the VGG team crawled data from more than 7000 speakers on YouTube in 2017~\cite{nagrani17_interspeech} and 2018~\cite{chung18b_interspeech} and performed automatic annotations. This dataset, called Voxceleb, has become the mainstream dataset for speaker recognition research in the past five years. Observing the results of the VoxCeleb Speaker Recognition Challenge (VoxSRC) in Figure \ref{fig:vox_res_trend}, the performance on the Voxceleb data is approaching saturation. Notably, although Voxceleb is `in the wild' data, its genre is relatively uniform, mostly consisting of interview data. Research in more challenging scenarios is still needed, such as in far-field~\cite{qin2020ffsvc,movsner2022multisv} and multiple genres~\cite{fan2020cn,li2022cn,lin2024voxblink,budaghyan2024cryceleb}. Additionally, the scaling law~\cite{kaplan2020scaling} has been validated; for NLP tasks, large datasets can enhance model performance. Whether the same is true for speaker recognition tasks remains to be studied. For this purpose, we have summarized all available speaker datasets in the Table \ref{table:dataset_list}. Furthermore, we have listed some atypical speaker datasets without speaker labels, such as Gigaspeech~\cite{chen21o_interspeech} and Wenetspeech~\cite{zhang2022wenetspeech}, to help researchers study the learning of speaker representation under less constrained conditions.

\subsection{Open-Source Toolkits}
\label{sec:tool}

Accessible and user-friendly open-source tools are pivotal for advancing research in the field of speaker recognition. Table~\ref{tab:toolkits} enumerates the relevant open-source toolkits. Kaldi, a speech processing toolkit that has remained popular for over a decade, provides numerous recipes for speaker recognition tasks. Additionally, many general-purpose speech platforms have integrated speaker recognition functionalities, including Espnet~\cite{jung2024espnet}, Speechbrain~\cite{ravanelli2021speechbrain}, and NeMo. Specialized open-source tools for speaker recognition, such as ASV-Subtools~\cite{tong2021asv}, VoxCeleb\_Trainer~\cite{nagrani2020voxceleb}, 3D-speaker~\cite{chen20243d}, and Wespeaker~\cite{wang2023wespeaker}, are also available. These toolkits are implemented in PyTorch and are easily modifiable, facilitating the extension of their frameworks by researchers.
Furthermore, self-supervised speaker recognition has garnered significant attention in recent years. Open-source frameworks like 3D-speaker~\cite{chen20243d} and Wespeaker~\cite{wang2024advancing} support such functionalities. To leverage the capabilities of large pretrained speech models, Espnet-Spk~\cite{jung2024espnet} and Wespeaker~\cite{wang2024advancing}  enable the integration of pretrained models such as WavLM and Whisper. To bridge the gap between research and real-world applications, Wespeaker and NeMo offer rapid deployment features, enabling swift implementation of models within industrial settings.

\begin{table}[!htb]
\caption{\label{tab:toolkits} Existing open-source toolkits which support deep speaker embedding learning}
\begin{adjustbox}{width=.5\textwidth,center}
\begin{tabular}{@{}lcccc@{}}
\toprule
Toolkit           & Speaker-task specific & SSL  & Pretrained Speech Models  & Deployment  \\ \midrule
Kaldi~\cite{povey2011kaldi}             & No                    & No                 & No        & No          \\\midrule
VoxCeleb\_Trainer~\cite{nagrani2020voxceleb} & Yes                   & No              & No     & No             \\\midrule
ASV-Subtools~\cite{tong2021asv}      & Yes                   & No                 & No     & Yes               \\\midrule
SpeechBrain~\cite{ravanelli2021speechbrain}       & No                    & No                 & No      & No              \\\midrule
NeMo\tablefootnote{\url{https://docs.nvidia.com/nemo-framework/user-guide/latest/nemotoolkit/asr/speaker_recognition/intro.html}}            & No                    & No        & No           & Yes                \\\midrule
Espnet~\cite{jung2024espnet}       & No                    & No      & Yes              & No                 \\\midrule
3D-Speaker~\cite{chen20243d}       & Yes                   & Yes          & No       & No                 \\\midrule
Wespeaker~\cite{wang2023wespeaker}         & Yes                   & Yes & Yes    & Yes                \\
\bottomrule  
\end{tabular}
\end{adjustbox}
\vspace{-5pt}
\end{table}

\section{Trends}
\label{sec:trends}

In this section, we discuss some of the challenges and future trends in speaker modeling. First, we reiterate several critical issues that require further exploration, including topics previously investigated:
1) The robustness and efficiency issues that significantly affect applications;
2) Effective methods to leverage large model pretraining;
3) Broadening the scope to other related tasks, seeking better adaptive speaker modeling methods;
4) Exploration of explainability.

Additionally, we would like to introduce several emerging trends that have not yet been covered.


\subsection{Privacy Protection and Ethical Issues}
\label{sec:privacy}

Since speaker modeling involves personal voice data and typically includes speaker identity labels, which may contain sensitive information, it is possible to infer a person's gender, age, and even pathological states from this data. Therefore, it is crucial to collect and use such Personally Identifiable Information (PII) without compromising personal privacy.
Although early datasets like RSR2015~\cite{larcher2014text} were collected with user consent, large-scale datasets commonly used today, such as VoxCeleb~\cite{nagrani2020voxceleb} and CNCeleb~\cite{li2022cn}, are often crawled from the web without obtaining the consent of the respective speakers. To address privacy concerns, techniques such as speaker anonymization~\cite{Tomashenko2021TheV2, srivastava2022privacy, Miao2023SpeakerAU, shamsabadi2022differentially, Meyer2022AnonymizingSW} can be employed to mask the original speaker's identity. The key challenge in this process is to ensure the quality of the anonymized data and its usability for downstream tasks. Researchers have also attempted to reconstruct anonymized datasets, such as SynVox2~\cite{Miao2023SynVox2TA}.

\subsection{Large Scale Self-supervised Representation Learning}
\label{sec:lssl}
Research from many other fields has already demonstrated that large-scale pretraining can enhance model performance, such as Whisper~\cite{radford2023robust} in speech recognition and BaseTTS~\cite{lajszczak2024base} in the TTS field. As we mentioned in the previous section, data with speaker labels involves privacy issues, which may prevent us from using large amounts of labeled data to learn speaker representations. Furthermore, the pretrained model introduced in section \ref{sec:pretrain} was not specifically designed for speaker representation learning during training. Additionally, the speaker self-supervised representation learning described in section \ref{sssec:self_supervised_method} has only been validated on datasets at the scale of Voxceleb. Moreover, the research~\cite{chen2023comprehensive} points out that the related algorithms do not bring performance improvements when scaled to larger datasets. Therefore, further research is needed if we want to use larger-scale unlabeled data for speaker representation learning.

\subsection{The Cross-modality Learning between Text and Speaker Representation}
\label{sec:cross}
Recently, large language models (LLMs)~\cite{achiam2023gpt,touvron2023llama,bai2023qwen} have garnered widespread attention from researchers. Beyond modeling text itself, many researchers have found that applying LLMs to various cross-modal fields can also be highly beneficial. In cross-modal tasks, LLMs~\cite{zhang2023speechgpt,islam2024gpt} are often used as an intermediary to connect different modalities. Similarly, it's crucial to explore how large language models can understand speaker representations, that is, converting speaker representations into textual descriptions. Conversely, it is also important to study how to convert textual descriptions of speaker characteristics into speaker representations. Regarding the understanding of speaker characteristics using text, ByteDance has proposed the SALMONN~\cite{chang_salmonn} model, which can use text to output speaker characteristics in speech. Similarly, there are other studies that generate speaker characteristics~\cite{guo2023prompttts,yang2024instructtts,zhang2023promptspeaker,chen24q_interspeech} based on text descriptions, but these studies are still in their early stages, and the performance is not very good. Further research is needed in the future.

\subsection{Attacks and Defenses}
\label{sec:con_attack}

Speaker representations are often utilized in identity verification systems, making the security of these systems crutial. With advancements in voice synthesis technology, zero-shot voice synthesis systems are now capable of accurately replicating a target person's voice using only a short reference audio sample. Furthermore, pre-trained speaker recognition systems frequently struggle to automatically differentiate between genuine and synthesized voices~\cite{kamble2020advances}. Additionally, neural network systems are inherently vulnerable; even without employing voice replication to deceive speaker systems, many attack algorithms can successfully compromise the system through slight perturbations~\cite{das2020attacker} of the input. Therefore, beyond the development of robust speaker recognition systems, there is a critical need to explore more powerful anti-spoofing and attack-resistant algorithms.

\subsection{Controlable Speaker Generation}
\label{sec:con_spkgen}
Currently, most speaker modeling primarily focuses on speech data from existing speakers, aiming to compress speaker information as completely, accurately, and efficiently as possible. However, in certain specific scenarios, we may need to model non-existent speaker identities, such as generating fictional voices in speech synthesis tasks or creating specific types of voices based on descriptions. This approach can generate voices with specific attributes that do not belong to any real speaker by finely adjusting speaker characteristics. This not only helps to protect user privacy but also provides a variety of voice options without infringing on the privacy of real speakers.
Existing research~\cite{wang2018style,valle2020flowtron,murata2024attribute} indicates that the simplest method is to perform linear interpolation on speaker embeddings. However, the ideal approach is to learn how to finely control speaker representations in a continuous space through different attributes. We firmly believe that this research direction holds great potential.

\subsection{Disentanglement Learning}
\label{sec:disentangle}
In the previous Section~\ref{sec:text}, we have already discussed some preliminary work on disentangling speaker content information. In recent years, the topic of disentanglement has gradually garnered growing attention. Liu et al.~\cite{liu2023disentangling} proposed a disentanglement framework that employs novel Gaussian inference layers with learnable transition models to capture speaker and content variability. Besides the benifits for text-independent speaker recognition, the effective disentanglement of speaker information and content information is also a crucial issue in the field of voice conversion and speech synthesis, especially in the zero-shot setups~\cite{ju2024naturalspeech}. With the recent emergence of Large Codec Language Models, some researchers have started to explore decoupling algorithms that are based on codecs~\cite{ju2024naturalspeech,zhang2023speechtokenizer}.

\section{Conclusion}
\label{sec:conclusion}
In this review article, from a unique perspective, we systematically review the speaker representation techniques based on deep learning, the evolution of algorithms, and their applications. We hope that this article provides a comprehensive and systematic summary, and offers detailed references and inspiration for researchers in the field of speaker modeling. Additionally, we aim to spark interest among researchers in related fields, and to promot the development and broader application of speaker modeling technologies. At the same time, we must acknowledge that due to space limitations and perspective of this paper, some aspects might only be briefly mentioned.

\bibliographystyle{IEEEtran}

\bibliography{mybib,reference/ssl,reference/dataset,reference/robustness,reference/old_era,reference/application,reference/analysis,reference/privacy}

\end{document}